\documentclass[10pt,journal,compsoc]{IEEEtran}

\ifCLASSOPTIONcompsoc
  \usepackage[nocompress]{cite}
\else
  \usepackage{cite}
\fi

\ifCLASSINFOpdf
\else
\fi

\usepackage{cite}
\usepackage{graphicx}
\usepackage{url}
\usepackage[ruled,linesnumbered]{algorithm2e}
\usepackage{amsmath}
\usepackage{booktabs}
\usepackage{multirow}
\usepackage{lscape}
\usepackage{xspace}
\usepackage[T1]{fontenc}
\usepackage[scaled=0.75]{beramono}
\usepackage{marvosym}
\usepackage{pifont}
\usepackage{float}
\usepackage{tikz}
\usepackage{comment}
\usepackage{color}
\usepackage{flushend}
\usepackage{balance}
\usepackage{array}
\usepackage{pdfpages}
\usepackage{fancyhdr}
\newcommand{\PreserveBackslash}[1]{\let\temp=\\#1\let\\=\temp}
\newcolumntype{C}[1]{>{\PreserveBackslash\centering}p{#1}}
\newcolumntype{R}[1]{>{\PreserveBackslash\raggedleft}p{#1}}
\newcolumntype{L}[1]{>{\PreserveBackslash\raggedright}p{#1}}

\newcommand*\circled[1]{\tikz[inner sep=0.5pt,baseline=-.75ex]{
            \node[shape=circle,draw] (char) {#1};}}
\newcommand{\tool}{\hbox{\textit{{BinSeeker}}}\xspace}
\newcommand{\toolm}{\hbox{\textit{{BinSeeker-}}}\xspace}
\newcommand{\gemini}{\hbox{\textit{{Gemini}}}\xspace}
\newcommand{\toolpre}{\hbox{\textit{{BinSeeker-}}}\xspace}
\newcommand{\Genius}{\hbox{\textit{{Genius}}}\xspace}
\newcommand{\Cacompare}{\hbox{\textit{{CACompare}}}\xspace}

\newenvironment{formula}{\begin{small}}{\end{small}}
\begin{document}

\title{Semantic Learning and Emulation Based Cross-platform Binary Vulnerability Seeker}

\author{Jian Gao, Yu Jiang, Zhe Liu, Xin Yang, Cong Wang, Xun Jiao, Zijiang Yang, Jiaguang Sun
\IEEEcompsocitemizethanks{\IEEEcompsocthanksitem J. Gao, Y. Jiang, X. Yang, C. Wang, and J. Sun are with the School of Software, Tsinghua University, Beijing National Research Center for Information Science and Technology, and Key Laboratory for Information System Security, Ministry of Education, Beijing 100084, China.\protect\\
E-mail: gaojian094@gmail.com, jiangyu198964@126.com, \{yangx16, wangcong15\}@mails.tsinghua.edu.cn, sunjg@tsinghua.edu.cn.
\IEEEcompsocthanksitem Z. Liu is with the College of Computer Science and Technology, Nanjing University of Aeronautics and Astronautics, Nanjing 211106, China. \protect\\
E-mail: zhe.liu@nuaa.edu.cn.
\IEEEcompsocthanksitem X. Jiao is with the Department of Electrical and Computer Engineering, Villanova University, Villanova, PA 19085 USA. E-mail: xujiao@eng.ucsd.edu.
\IEEEcompsocthanksitem Z. Yang is with the Department of Computer Science, Western Michigan University, Kalamazoo, MI 49008 USA. E-mail: zijiang.yang@wmich.edu.
}
\thanks{Manuscript received 6 Feb. 2019; revised 29 Oct. 2019; accepted 16 Nov. 2019. Date of publication 2 Dec. 2019; date of current version 12 Nov. 2021.}
\thanks{(Corresponding author: Yu Jiang.)}
\thanks{Recommended for acceptance by X. Zhang.}
\thanks{Digital Object Identifier no. 10.1109/TSE.2019.2956932}
}

%
%

\IEEEtitleabstractindextext{%
\begin{abstract}
Clone detection is widely exploited for software vulnerability search. The approaches based on source code analysis cannot be applied to binary clone detection because the same source code can produce significantly different binaries due to different operating systems, microprocessor architectures and compilers.
In this paper, we present \tool, a cross-platform binary seeker that integrates semantic learning and emulation.
With the help of the labeled semantic flow graph, \tool can quickly identify $M$ candidate functions that are most similar to the vulnerability from the target binary. The value of $M$ is relatively large so this semantic learning procedure essentially eliminates those functions that are very unlikely to have the vulnerability.
Then, semantic emulation is conducted on these $M$ candidates to obtain their dynamic signature sequences. By comparing signature sequences, \tool produces top-$N$ functions that exhibit most similar behavior to that of the vulnerability.
With fast filtering of semantic learning and accurate comparison of semantic emulation, \tool seeks vulnerability precisely with little overhead.
The experiments on six widely used programs with fifteen known CVE vulnerabilities demonstrate that \tool outperforms three state-of-the-art tools \Genius, \gemini and \Cacompare.
Regarding search accuracy, \tool achieves an MRR value of 0.65 in the target programs, whereas the MRR values by \Genius, \gemini and \Cacompare are 0.17, 0.07 and 0.42, respectively.
If we consider ranking a function with the targeted vulnerability in the top-5 as accurate, \tool achieves the accuracy of 93.33 percent, while the accuracy of the other three tools is merely 33.33, 13.33 and 53.33 percent, respectively. Such accuracy is achieved with 0.27s on average to determine whether the target binary function contains a known vulnerability, and the time for the other three tools are 1.57s, 0.15s and 0.98s, respectively. Compared to the time used to manually identify the true positive vulnerability from the false positive candidates reported by Gemini, the time overhead of \tool is negligible.
Evidently, the proposed \tool achieves a better balance between accuracy and efficiency.
\end{abstract}

}

\maketitle

\thispagestyle{fancy}

\lfoot{\footnotesize 0098-5589 © 2019 IEEE. Personal use is permitted, but republication/redistribution requires IEEE permission. See https://www.ieee.org/publications/rights/index.html for more information.}

\cfoot{}

\renewcommand{\headrulewidth}{0mm}

\IEEEdisplaynontitleabstractindextext

%
\IEEEpeerreviewmaketitle

\section{Introduction}
\label{introduction}
\IEEEPARstart{P}{re-existing} code has been widely reused to improve software development productivity.
Releases of different software products contain significant amounts of identical or similar code, a phenomenon known as code clone.
The related study has shown that 22.3\% of the Linux kernel is from the previous implementation \cite{cp_miner}.
Consequently, along with the widespread usage of code clone, vulnerabilities spread as well because the code fragments containing the vulnerabilities are reused in other software products.
In addition, the patches that fix vulnerabilities are not automatically propagated, thus the cloned vulnerabilities are not patched even though the original code has been fixed.
For example, 145 unpatched cloned vulnerabilities were confirmed in the Debian system \cite{JangAB12}.

In order to maintain code quality, it is critical to identify cloned code once a vulnerability is detected.
Unfortunately, it is a very challenging task because the same source code can be compiled with different compilation parameters and even different compilers.
In addition, with the popularity of terminal devices, software programs on traditional \textit{X86} architecture are gradually ported to other architectures such as \textit{ARM} and \textit{MIPS}.
For example, the cyber-physical systems \cite{Design,Design-mixed,Data-centered,Safety-assured} for different scenarios may contain different binaries compiled from the same source code.
To ensure that these security-critical systems are protected from the same vulnerability, it is especially important to search for known vulnerabilities from binaries with different architectures.
To address this challenge, cross-platform binary vulnerability search has attracted increasing attention in recent years, and three types of approaches have been proposed: static, dynamic and learning approaches~\cite{discovRE, Bingo, CACOMPARE, CCS17-Xu, vulseeker}.

Static approaches usually rely on the graph matching algorithm on control flow graphs (CFGs) to identify binary code similar to the vulnerable code \cite{Luo2014SemanticsbasedOB, DBLP:conf/sp/PewnyGGRH15, discovRE, Bingo}.
However, CFGs of the same function differ significantly with different compilation configurations (e.g., \textit{O0{\textendash}O3} optimization levels, \textit{X86} and \textit{MIPS} processor architectures), which leads to inaccurate results.
In order to enhance accuracy, Pewny \textit{et al.} \cite{DBLP:conf/sp/PewnyGGRH15} presents a method to broaden the analysis from the similarity of basic blocks to the similarity of functions based on CFGs.
Although this method improves accuracy, having too many false positives prevents it from being widely adopted.

Dynamic approaches overcome the obstacle of inaccurate search through monitoring the runtime traces of binary programs in real operating environments and then performing equivalence checking between two traces \cite{Ming2017BinSimTS, CACOMPARE}.
However, significant overhead is inherent and dynamic approaches are often expensive in practice.
For example, although \Cacompare \cite{CACOMPARE} achieves over 80\% search accuracy, it takes \Cacompare about three hours to generate function signatures and about seven hours to complete the search task for the \textit{OpenSSL} with 5,995 functions.
Though fuzzy testing techniques \cite{afl, aflfast, enfuzz, PAFL, driller} are also effective vulnerability search approaches, their performance is limited for cross-platform binaries.

As a burgeoning technique, learning-based approaches are increasingly applied to binary vulnerability search because of fewer domain knowledge requirements \cite{CCS16-Feng,CCS17-Xu, vulseeker}.
Since most of these approaches rely only on the CFGs to convert assembly instruction features into numerical vectors, they can quickly predict whether a binary function is vulnerable.
Unfortunately, their accuracy is rather low.
For example, the top-50 accuracy of \Genius for two case studies on a large set of firmware images is only 28\% and 48\% \cite{CCS16-Feng}.
With such a low accuracy, nontrivial manual effort is often required to eliminate false positives, which is not practical for industrial applications.
To enhance the robustness against structural differences in the CFG, we propose \textit{VulSeeker} \cite{vulseeker}, a semantic learning-based approach that adopts lightweight instruction features and integrates DFG into CFG.
Furthermore, we also attempt to supplement multiple semantic signatures in \textit{VulSeeker-Pro} \cite{vulseeker-pro} to evaluate function similarity against compilation optimization differences on the same architecture.

Both learning-based and dynamic approaches have barriers to use when it comes to large-scale code.
For example, consider such a scenario to determine whether a large number of cross-architecture embedded firmware contain a serious vulnerable function that has just been exposed.
In this case, applying the learning-based approach alone can quickly predict the similarity between any function within the firmware and the vulnerable function.
Since it only uses vectors to represent the cross-architecture function semantics which can be insufficient, additional human efforts are required to participate in the confirmation process.
On the other hand, the dynamic approach can well match the same vulnerable function of different platforms, but may suffer from time-demand bottlenecks of large-scale code. Execution for all functions is extremely time consuming.

To overcome the above limitations and strike a balance between accuracy and efficiency, we present \tool\footnote {The prototype implementation is open-source and is available at {\url{https://github.com/buptsseGJ/BinSeeker.}}},
a cross-platform binary vulnerability seeker that integrates learning and dynamic approaches.
It seamlessly integrates our previous works  \cite{vulseeker,vulseeker-pro} and supports the cross-architecture function emulation to achieve better performance. The intuitive idea is to apply the dynamic approach to the quickly acquired narrow range of candidate functions that are similar to the vulnerability, which can both improve the search accuracy and reduce the time requirement. Similar idea has also been applied in Driller for vulnerability fuzzing, that combine the cheap searched based AFL to achieve fast coverage and the expensive symbolic execution to solve the constraints hard to be searched.

Given a vulnerable function, the learning component in \tool quickly identifies $M$ candidate functions from the target binary that are most similar to the vulnerable function.
With the aid of the labeled semantic flow graph,  \tool captures more function-level semantics and is more accurate than the existing learning-based approach \cite{CCS17-Xu}.
However, $M$ has to be relatively large due to the inherent inaccuracy in all learning-based approaches; otherwise, false negatives are very likely to occur.
To avoid manual efforts of examining all $M$ functions, the semantic emulation component of \tool conducts dynamic emulation on the $M$ candidates and identifies top-$N$  functions that are most similar to the vulnerable function. That is,
\tool first conducts a computationally cheap and less accurate search on the entire target program using semantic learning, and then performs an expensive and more accurate search on the $M$ candidate functions using semantic emulation.
With  $N$ being much smaller than $M$, \tool obtains highly accurate results at a cost almost the same as that of a semantic learning-based approach.

To evaluate \tool on widely-used open-source applications such as \textit{OpenSSL}, we compare it with three state-of-the-art cross-platform binary vulnerability search tools: \Genius~\cite{CCS16-Feng} and \gemini~\cite{CCS17-Xu} which are semantic learning-based tools and \Cacompare~\cite{CACOMPARE} which is a semantic emulation-based tool.
The experimental results show that \tool significantly outperforms the three tools in comparison.
{\color{black}Regarding vulnerability search accuracy, \tool achieves an MRR (mean reciprocal rank) value of 0.65 in the target programs, whereas the MRR values of \Genius, \gemini and \Cacompare are 0.17, 0.07 and 0.42, respectively.
If we consider ranking a function with the targeted vulnerability in the top-5 as accurate, \tool achieves a top-5 accuracy of 93.33\%, while the top-5 accuracy of the other three tools is merely 33.33\%, 13.33\% and 53.33\%, respectively.}
In terms of time used in the search, it takes \Genius, \gemini, \Cacompare and \tool 8,992s, 849s, 8,432s and 1,323s respectively to complete a search task on the \textit{OpenSSL} binary.
The results demonstrate that \tool is more accurate than semantic emulation-based tools and the cost is comparable to that of the learning-based approach.
These advantages reduce the burden of engineers when manually identifying true positive cases.
In summary, the present study makes the following contributions:
\begin{itemize}
\item To date, \tool is the first tool that integrates semantic learning and emulation to improve search accuracy and efficiency. Accuracy is critical as it reduces manual efforts to inspect a large number of functions.
Meanwhile, \tool achieves highly accurate results at a cost almost the same as that of the fast semantic learning-based approach.

\item {\color{black}We optimize the original learning and emulation approaches to improve \tool's performance. Consequently, the learning component is more accurate with the extended labeled semantic flow graph and the emulation engine is more lightweight and faster with the optimized function signature extraction.}

\end{itemize}

The rest of this paper is organized as follows:
Section \ref{background} introduces the background to help understand our neural network model;
Section \ref{design} details the design and implementation of \tool;
Section \ref{experiment} describes experimental results compared to the state-of-the-art approaches;
and Section \ref{threat-to-validity} discusses challenges and future work.
Section \ref{related-work} delineates related work and Section \ref{conclusion} presents the conclusion.

\section{Background}
\label{background}
\begin{figure}[!htbp]
    \centering
    \includegraphics[width=3.5in]{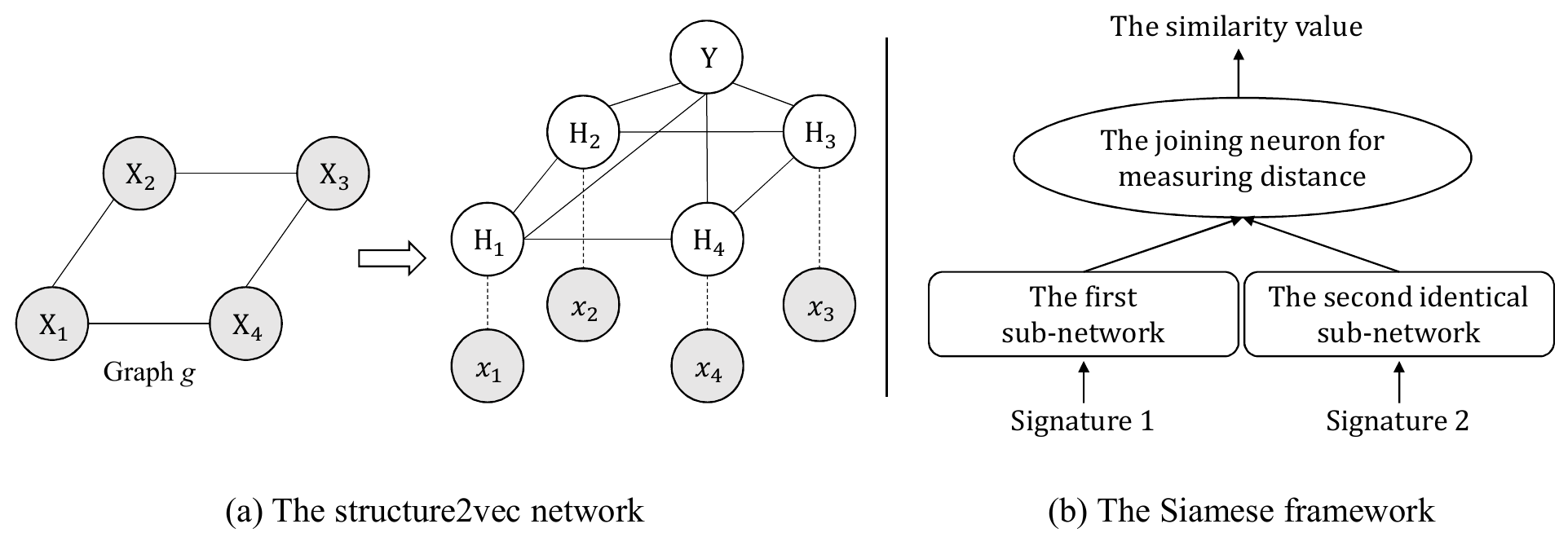}
    \caption{\color{black}Two foundational network structures used in \tool.}
    \label{fig-structure2vec}
\end{figure}

The semantic learning module of \tool is based on the \textit{Siamese} framework \cite{siamese} and the \textit{structure2vec} network \cite{struct2vec}, as shown in Fig. \ref{fig-structure2vec}.

\textbf{a) Application of the \textit{Siamese} Framework and  the \textit{structure2vec} Network to Vulnerability Search.}  In order to identify the functions that are most similar to the one with known vulnerability,  a key step is to compute the similarity between a pair of functions.
In \tool, the input pair of functions are represented as LSFGs (described in Section \ref{LSFG}), and the neural networks are dedicated to the LSFG structures.
The embedding vector generation strategy of the \textit{structure2vec} network transforms the vulnerability search problem to be the problem of calculating the similarity between function embedding vectors.
Given two graphs $g_1$ and $g_2$, the label $y$ in the training tuple $\left \langle g_1,g_2,y \right \rangle$ indicates whether the two graphs are similar.
This dual-input and single-output tuple requires us to train two \textit{structure2vec} networks, one for each graph,  with shared parameters.
{\color{black}Thus, we embed two identical \textit{structure2vec} networks into the \textit{Siamese} framework shown in Fig. \ref{fig-network}(a), which is described in detail in Section \ref{approach: DNN}.
The final effect is that the vector representations of similar graphs are close to each other.}

\textbf{b) The \textit{structure2vec} Network.}
Dai \textit{et al.} \cite{struct2vec} proposed the \textit{structure2vec} graph neural network, which proved to be effective and scalable for structured data representation.
The vertices in the graph are connected to each other by the edges.
The \textit{structure2vec} network can encode vertex features and connection relationship of the edges in the graph as the embedding vector to represent graph semantics.
Fig. \ref{fig-structure2vec}(a) shows how the \textit{structure2vec} network works to generate the embedding vector based on the graph $g$ on the left.
The example graph contains four vertices $X_{i}$, $i \in \left \{ 1,2,3,4 \right \}$, and each vertex is appended with an initial numerical vector $x_{i}$.
The right side of Fig. \ref{fig-structure2vec}(a) is the  \textit{structure2vec} network structure that consists of one input layer, $T$ hidden layers (only one hidden layer is drawn for illustration), and one output layer.
In the \textit{structure2vec} network, the number of neuron nodes in the input layer and each hidden layer is the number of vertices in the original graph $g$.

\begin{figure*}[!htbp]
    \centering
    \includegraphics[width=6.9in, height=1in]{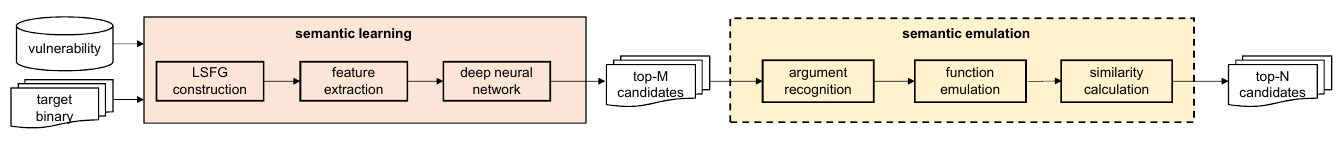}
    \caption{Overall workflow of \tool: the first phase relies on semantic learning to quickly predict top-M most similar candidate functions with a low time cost, the second phase employs function emulation to output more accurate top-N candidates, where M and N can be dynamically configured.}
    \label{fig-overall}
\end{figure*}

In the input layer, the input of each neuron node is the initial numerical vector $x_{i}$ of the corresponding vertex.
In the $t^{th}$ ($1 \leq t\leq T$) hidden layer, each hidden neuron $H_{i}$ is responsible for generating a new feature representation called the embedding vector $\widetilde{\mu }_{i}^{(t)}$.
We use formula $\widetilde{\mu}_{i}^{(t)}=F(x _{i}, \Sigma_{k\in E(i)}\widetilde{\mu}_{k}^{(t-1)})$ to represent the mapping relationship of each hidden neuron, where $E(i)$ refers to the set of vertices adjacent to vertex $X_{i}$.
Through the mapping function $F$, the feature of each vertex is propagated to other vertices based on connected edges.
This feature update strategy takes into account the graph topology and ensures that each vertex incorporates information from neighbors within $T$ hops after $T$ hidden-layer iterations.
Finally, the output layer neuron $Y$ aggregates the output embedding vectors $\widetilde{\mu}_{i}^{(T)}$ of the $T ^{th}$ hidden layer neurons to form the final embedding vector of the entire graph.

\textbf{c) The \textit{Siamese} Framework.}
The \textit{Siamese} framework \cite{siamese} shown in Fig. \ref{fig-structure2vec}(b) solves the dual input problem of verification of signatures written on a pen-input tablet.
It mainly consists of two identical sub-networks, each of which takes a processed signature as its input and then outputs the feature of the signature.
The joining neuron is used to measure the distance between these two output features of the two sub-networks and output the similarity value ranging from -1 to 1.
In addition, it is used as a network framework to train two identical embedded sub-networks in an end-to-end manner, which ensures that the two sub-networks share the same parameters.

\section{Design of \tool}
\label{design}

Our objective is to automatically identify whether a given binary program contains known vulnerabilities or not.
Lots of vulnerability-related databases are open to the public.
For example, \textit{Common Vulnerabilities and Exposures} (CVE) \cite{CVE} contains more than 120,000 vulnerabilities.
In the case of open-source software, which is often reused to improve software development, some of the CVEs provide an explicit indication of the source functions involved in the vulnerabilities.
Based on the indication of these known vulnerabilities within a single function, clone-based search approaches at the function-level granularity will help to detect those known vulnerabilities in target cross-platform binaries.

As presented in Fig. \ref{fig-overall}, \tool contains two major modules: \textit{semantic learning} and \textit{semantic emulation}.
Its inputs are a specific vulnerability and a target binary to be searched.
\tool utilizes the fast predictive capability of the customized deep neural network to obtain the initial $M$ (e.g., 200) candidate functions by eliminating extremely dissimilar functions in the \textit{semantic learning} module.
Then, \tool sorts the $M$ candidate functions based on function dynamic signature sequences to generate top-$N$ (e.g., top-25) candidate functions as the final prediction results for the vulnerability in the \textit{semantic emulation} module.

\subsection{Semantic Learning}
The main goal of this module is to quickly eliminate remarkably dissimilar functions from the target binary and get the top-M candidate functions that are most similar to a specific vulnerability.
Its pivotal capability is to obtain embedding vectors representing the function semantics and use them for similarity comparison according to three steps: LSFG construction, feature extraction, and customized deep neural network.

\subsubsection{LSFG Construction}
\label{LSFG}
Many existing methods rely on CFGs as a basis to obtain function semantic representations.
These representations are either the symbolic formulas encoding the input-output relationship of basic blocks \cite{Lakhotia2013FastLO, DBLP:conf/sp/PewnyGGRH15, Luo2014SemanticsbasedOB} or the attributed control flow graphs containing instruction features \cite{CCS16-Feng,CCS17-Xu}.
Function semantics obtained through these methods can be highly inaccurate since the CFGs show significant diversities under different compilation scenarios \cite{tracelet}.

We propose the labeled semantic flow graph (LSFG) which combines the CFG and the data flow graph (DFG) to capture more accurate function semantics.
The idea is based on the fact that the CFG determines the possible execution sequences of basic blocks and the DFG depicts the transfer and use of data within the function.
The combination of these two dependent relations makes function semantics (explained in Section \ref{approach: DNN}) resistant to structural and syntactic differences in the CFGs under different architectures and compilation optimization strategies.

LSFG is different from the program dependence graph (PDG) \cite{PDG} and hybrid information- and control-flow graph (HI-CFG) \cite{HI-CFG} which also consists of DFG and CFG.
They work at different code granularity as well.
For example, PDG establishes an edge connection on the statement or instruction granularity, while LSFG is on the basic block granularity.
HI-CFG not only needs to create data structure nodes but also code block nodes.
It also needs to use the trace-based dynamic analysis approach to infer the edge connections between these two types of nodes.
Considering that the complexity of the HI-CFG and PDG structures sharply increases the processing time of the semantic learning model, we choose the more lightweight LSFG graph representation which only creates code block nodes in the CFG but ignores the data dependence edges within each basic block of the CFG.
In terms of the efficiency and accuracy of the experiment, the proposed LSFG is suitable for the customized structure2vec-based semantic learning network to learn the numerical semantic representation of the binary function.

\begin{figure}[!htbp]
    \centering
    \includegraphics[width=3.5in]{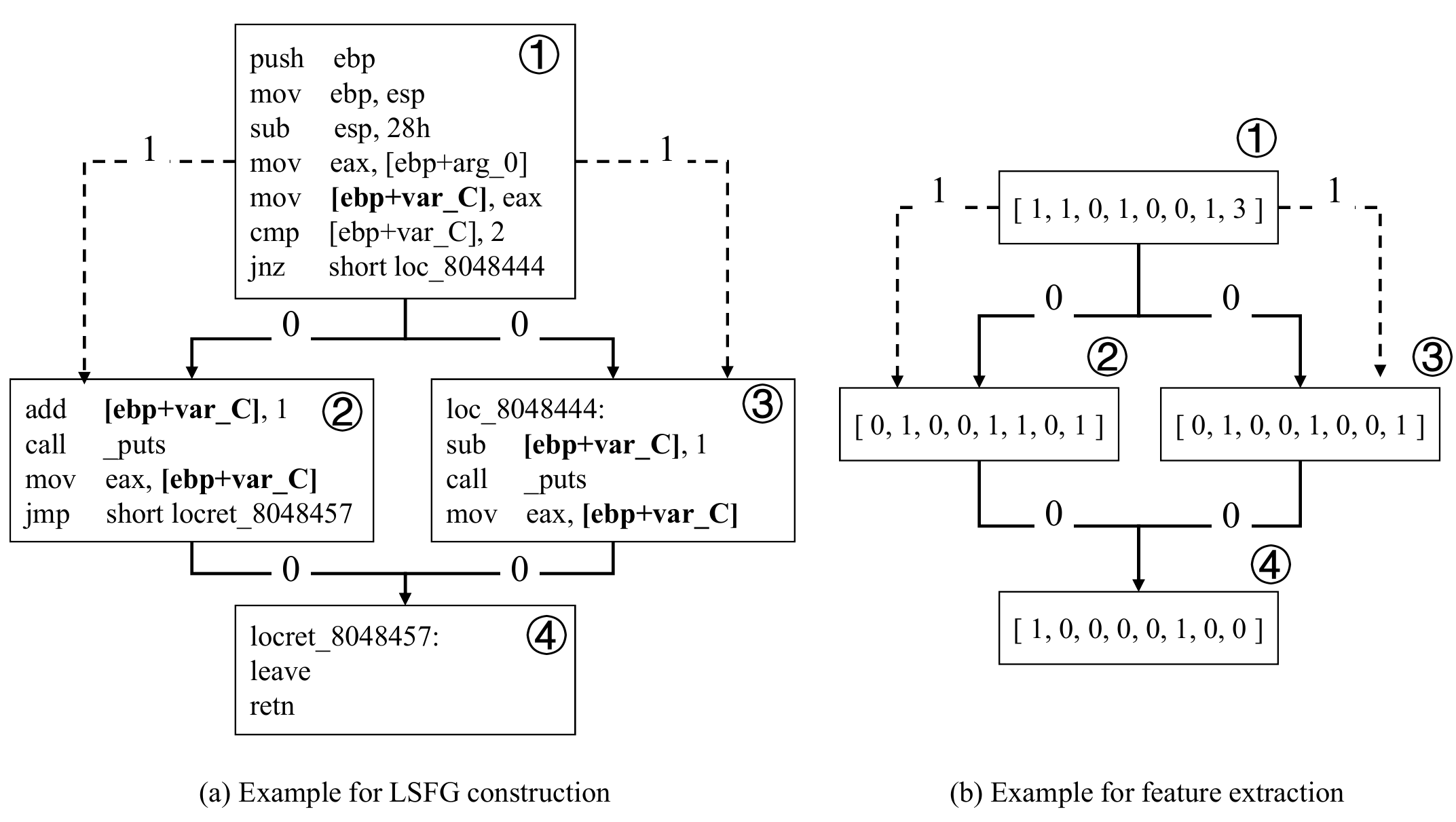}
    \caption{An example for the deep neural network input vector generation.}
    \label{fig-LSFG}
\end{figure}

Fig. \ref{fig-LSFG}(a) illustrates an example of the proposed LSFG.
The solid lines (labeled by 0) represent control dependency, while the dotted lines (labeled by 1)  represent data dependency.
For a compiled binary function, the structure of its assembly function is organized according to the CFG, which can be easily parsed with the help of common disassembly software (e.g., \textit{IDA Pro} \cite{IDA}, \textit{angr} \cite{angr}).
CFGs obtained by different methods are almost the same, while DFGs differ according to different data dependence rules.
In the present paper, we construct the DFGs on top of CFGs by leveraging the \textit{define-use} rules and traversing all function paths.
Specifically, for two instructions $i$ and $j$ from two different basic blocks, if the instruction $i$ writes to a memory location and the instruction $j$ reads from the same memory location,
we create a data dependence edge from the basic block of $i$ to the basic block of $j$.
It is worth noting that there is at most one data dependence edge between two different basic blocks.
When a variable is directly read without being written in the current basic block, we look for the pre-ordered basic block in which the variable is last defined, and then establish the data pointing edge between the two basic blocks unless the variable is never defined in the function.
Therefore, the presence of memory address ``\textit{$[ebp+var\_C]$}'' forms a data dependence edge between the block \circled{1} and \circled{2} in Fig. \ref{fig-LSFG}(a).

\begin{figure*}[t]
    \centering
    \includegraphics[width=6.3in]{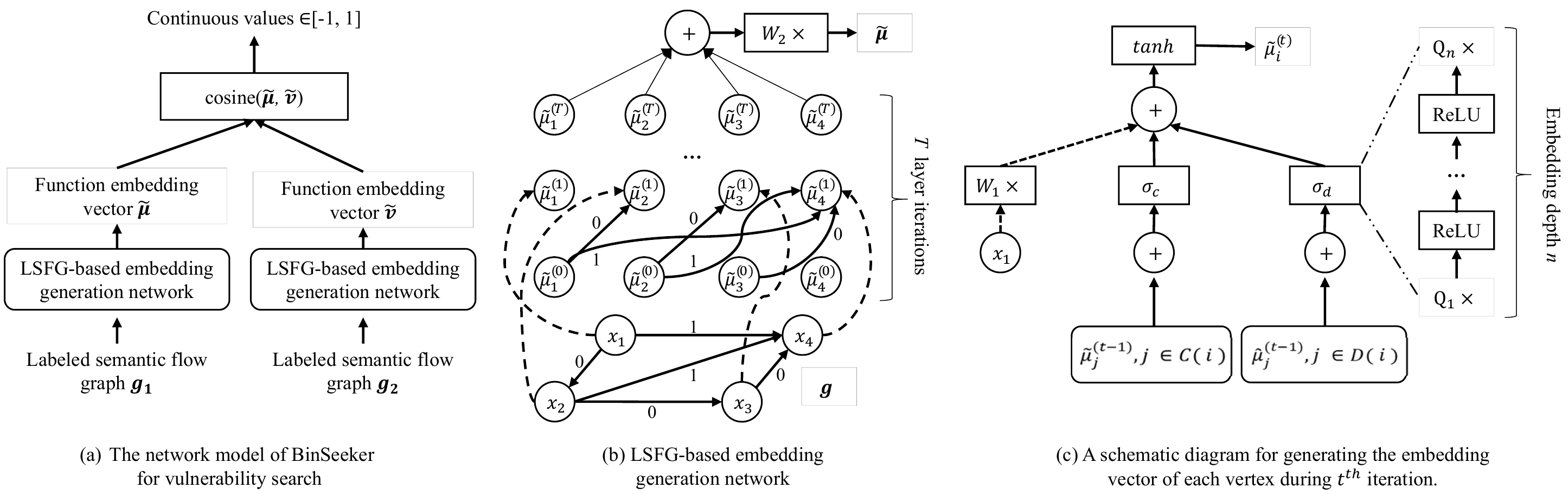}
    \caption{\color{black}The network of \tool for vulnerability search, including the customized similarity calculation, embedding and vector generation.}
    \label{fig-network}
\end{figure*}

\subsubsection{Feature Extraction}
\label{sec:features}
The LSFG constructed above is not suitable yet as input into our customized neural network (NN) model.
The purpose of feature extraction is to generate the  block-level initial numerical vectors related to functions that can be input into the NN model to generate function-level embedding vectors for similarity calculation.
We should choose and extract  robust and lightweight features that change little under various implementation platforms with different microprocessor architectures and various compilation optimization configurations as initial numerical vectors.
By empirically referring to features used in previous works \cite{CCS16-Feng,discovRE, alrabaee2016bingold} and executing a series of experiments (described in Section \ref{experiments:LSFG_feature}) for different feature sets, we finally determine to use the 8 types of features shown in Table \ref{table-features}.

\begin{table}[!htbp]
\centering
\caption{Basic-block level features used by \tool.}
\label{table-features}
\small
\begin{tabular}{@{}L{6.0cm}L{2.0cm}}
\toprule
\textbf{Feature Name}                           & \textbf{Example}      \\
\midrule
No. of stack operation instructions    & push, pop    \\
No. of arithmetic instructions         & add, sub     \\
No. of logical instructions            & and, or      \\
No. of comparative instructions        & test         \\
No. of library function calls          & call printf  \\
No. of unconditional jump instructions & jmp          \\
No. of conditional jump instructions   & jne, jb      \\
No. of generic instructions            & mov, lea     \\
\bottomrule
\end{tabular}
\end{table}

We first count the number of each feature in each basic block, then arrange them into numerical vectors in order,
and finally put these numerical vectors at the corresponding vertex of LSFG.
Fig. \ref{fig-LSFG}(b) presents the numerical  vectors of each basic block corresponding to the function in Fig. \ref{fig-LSFG}(a).
We denote the LSFG with numerical vectors as $g=\left\langle X,C,D \right\rangle$, where $X$, $C$ and $D$ are the sets of (basic block) vertices, control dependence edges and data dependence edges, respectively.
Each vertex $x_{i} \in X$ represents the initial numerical feature vector.
The LSFG mentioned later in this paper refers to the LSFG with initial numerical feature vectors unless otherwise specified.
Paired LSFGs $g_1$ and $g_2$ are the input of the \tool neural network.

\subsubsection{Customized Deep Neural Network}
\label{approach: DNN}
Because the two basic networks introduced in Section \ref{background} satisfy our vulnerability search requirement, this part clarifies how to combine the \textit{Siamese} \cite{siamese} framework with the adapted \textit{structure2vec} \cite{struct2vec} network to implement our customized network.
Since the \textit{structure2vec} network is implemented to support LSFGs, we call it the LSFG-based embedding generation network.

\textbf{a) \textit{BinSeeker} Network Structure.} Fig. \ref{fig-network}(a) shows the overall architecture of the \tool neural network model.
Its inputs are two LSFGs, which are abbreviated as $g_{1}$ and $g_{2}$.
These two graphs ($g_{1}$ and $g_{2}$) are imported into two identical LSFG-based embedding generation networks,  
which transform the structured graph information into function-level embedding vectors (e.g., $\widetilde{\mu }$, $\widetilde{\nu }$) capturing the function semantics.
The \textit{Cosine} function is used to calculate the similarity of two embedding vectors, which represents the similarity of two binary functions.

Fig. \ref{fig-network}(b) is a detailed description of the LSFG-based embedding generation network.
This process is similar to the basic \textit{structure2vec} except for the graph type.
We extend \textit{structure2vec} to deal with LSFG containing both data flow (labeled 1 on the solid arrow) and control flow (labeled 0 on the solid arrow), while basic \textit{structure2vec} works with an ordinary undirected graph.
In the input layer, the input example LSFG $g=\left\langle X,C,D \right\rangle$ consists of four vertices $X_{i}$, $i \in \left \{1,2,3,4  \right \}$, each of which represents a basic block of the function and contains the block-level initial feature vector $x_{i}$.
$C(i)$ and $D(i)$ represent the control dependence edge set and the data dependence edge set of vertex $i$, respectively.

The adapted \textit{structure2vec} network contains a total of $T$ hidden layer responsible for transforming graph information into the function semantic embedding vector.
Each hidden layer node is represented as the updated block-level embedding vector $\widetilde{\mu} _{i}^{(t)}$, where different values of $t$ correspond to different hidden layers.
During the $t^{th}$ hidden layer iteration, the updated $\widetilde{\mu} _{i}^{(t)}$ consists of three different inputs: the initial feature vector $x_{i}$ of the corresponding vertex $X_{i}$ (the dotted arrow in Fig. \ref{fig-network}),
the sum $l_{c}^{(t-1)}=\Sigma _{j \in \mathit{C(i)}}\widetilde{\mu} _{j}^{(t-1)}$ of previous embedding vectors of vertices pointing to $X_{i}$ through the control dependency $C(i)$, and the sum $l_{d}^{(t-1)}=\Sigma _{j \in \mathit{D(i)}}\widetilde{\mu} _{j}^{(t-1)}$ of previous embedding vectors of vertices pointing to $X_{i}$ through the data dependency $D(i)$.
The updated $\widetilde{\mu} _{i}^{t}$ is expressed by mapping function $F$ with the formula: $\widetilde{\mu} _{i}^{(t)} = F(x_{i},l_{c}^{(t-1)},l_{d}^{(t-1)}) = tanh (\mathit{W}_{1}x_{i} + \sigma _{c} (l_{c}^{(t-1)}) + \sigma _{d} (l_{d}^{(t-1)}))$.
Fig. \ref{fig-network}(c) illustrates the procedure for generating the embedding vector $\widetilde{\mu} _{i}^{(t)}$ of each hidden node, where $\sigma _{c}$, $\sigma _{d}$ are two non-linear transformation functions which are responsible for calculating an embedding vector with more powerful representation capability.
Similar to \cite{CCS17-Xu}, we define them as two $n$ layer fully-connected networks with the following equations:

{\begin{small}
\begin{equation}
\left\{\begin{matrix}
\sigma _{c}(l _{c}) = P_{1}\times ReLU(P_{2}\times \cdots ReLU(P_{n} \times l_{c}))\\
\sigma _{d}(l _{d}) = Q_{1}\times ReLU(Q_{2}\times \cdots ReLU(Q_{n} \times l_{d}))
\end{matrix}\right.
\end{equation}
\end{small}}{where $n$ is the embedding depth representing the number of layers in a fully-connected network, $P_{i}$ and $Q_{i}$ are $p \times p$ dimensional parameter matrices for each layer of the two fully-connected networks.}

The overall procedure for generating function semantic representation described above is integrated into the Algorithm \ref{alg-network}.
$W_{1}$ and $W_{2}$ are $d \times p$ and $p \times p$ dimensional parameter matrices, respectively.
Through the $T$-layer iteration from Lines 5 to 11 in Algorithm \ref{alg-network}, a new embedding vector of each vertex is generated, which not only follows the topology structure of LSFG but also integrates the $T$-hop interaction among vertices.
In other words, the features of the vertices are propagated to other vertices as the iteration progresses, ensuring that each basic block within the function incorporates information from neighbors within T hops after T hidden-layer iterations. Finally, the binary function semantics, including the data flow dependency and the control flow dependency, is aggregated into the function-level embedding vector $\widetilde{\mu}$ in Line 12.

\begin{small}
\begin{algorithm}
\small
    \caption{Generating function semantics}
    \label{alg-network}
    \KwIn{LSFG $g = \left\langle X,C,D\right\rangle$\\\quad \quad \quad Hidden layer iteration number $T$}
    \KwOut{Binary function semantics embedding vector $\widetilde{\mu }$}
    \begin{small}
    $C(i)$ as the set of parent nodes that are the control dependency of vertex $i$;
    $D(i)$ as the set of parent nodes that are the data dependency of vertex $i$.
    \end{small}
    \\
    \For{$i \in X$}
    {
        $\widetilde{\mu} _{i}^{(0)} = 0$
    }
    \For{$t = 1$ to $T$}
    {
        \For{$i \in X$}
        {
            $l_{c}^{t-1} = \Sigma _{j \in \mathit{C(i)}}\widetilde{\mu} _{j}^{(t-1)}$\\
            $l_{d}^{t-1} = \Sigma _{j \in \mathit{D(i)}}\widetilde{\mu} _{j}^{(t-1)}$\\
            $\widetilde{\mu _{i}}^{(t)}=tanh (\mathit{W}_{1}x_{i} + \sigma _{c} (l_{c}^{t-1}) + \sigma _{d} (l_{d}^{t-1}))$
        }
    }
    return $\widetilde{\mu} = W_{2}(\Sigma _{i \in X}\widetilde{\mu} _{i}^{(T)})$
\end{algorithm}
\end{small}

\textbf{b) Learning Parameters.}
Paired embedding vectors (e.g., $\widetilde{\mu}$, $\widetilde{\nu}$) are obtained through two identical LSFG-based embedding generation networks with two LSFGs (e.g., $g_{1}$ and $g_{2}$) as inputs.
The output of the whole network represents the similarity of the two functions and is measured by the \textit{Cosine} function denoted as $\widehat{y} = cos(\widetilde{u},\widetilde{v}) = (\widetilde{u} \cdot \widetilde{v})/(\begin{Vmatrix}\widetilde{u}\end{Vmatrix} \cdot \begin{Vmatrix}\widetilde{v}\end{Vmatrix})$, where $\widehat{y}$ is the predicted similarity output of two functions, ranging from $-1$ to $1$.
Given the ground truth $y\in \left \{ 1,-1 \right \}$ of LSFGs $g_{1}$ and $g_{2}$, $y=1$ indicates they are similar functions; otherwise, they are dissimilar.
Suppose that the training data set has $M$ pairs of labeled samples $\left \langle g_{1},g_{2},y \right \rangle$, then our training objective is to minimize the training errors.
We use the stochastic gradient descent algorithm to minimize the error function $min \, E(W_{1},W_{2},P_{1}\cdots P{n},Q_{1}\cdots Q_{n})=\sum_{m=1}^{M}(\widehat{y}-y)^{2}$ and obtain the most appropriate model parameters (e.g., $W_{1}, P_{1}$).

\subsection{Semantic Emulation}
The primary goal of semantic emulation is to sort the $M$ candidate functions obtained from the semantic learning module. By exploiting three steps to obtain  more accurate results,  semantic emulation reduces the burden of engineers when manually inspecting the $M$ functions: argument recognition, function emulation, and similarity calculation.

\subsubsection{Argument Recognition}
Before emulating the execution of a function, \tool first needs to recognize the required function arguments, which are usually classified as register arguments and stack arguments.
With the aid of \textit{IDA Pro} \cite{IDA}, argument recognition is implemented based on the disassembled assembly code from the binary program.
For example, the first three arguments of a function may be register arguments stored in the \textit{EAX}, \textit{EDX} and \textit{ECX} registers in the \textit{X86} architecture.
The remaining arguments are passed through the program stack, whose space grows from the high address to the low address.
Each function has a stack pointer indicating the stack start address.
When traversing the assembly instructions, if an instruction accesses a stack address that is larger than the stack start address, the offset of the address relative to the stack start address is recorded as a stack argument.

\subsubsection{Function Emulation}
\begin{figure}[!htbp]
    \centering
    \includegraphics[width=3.4in]{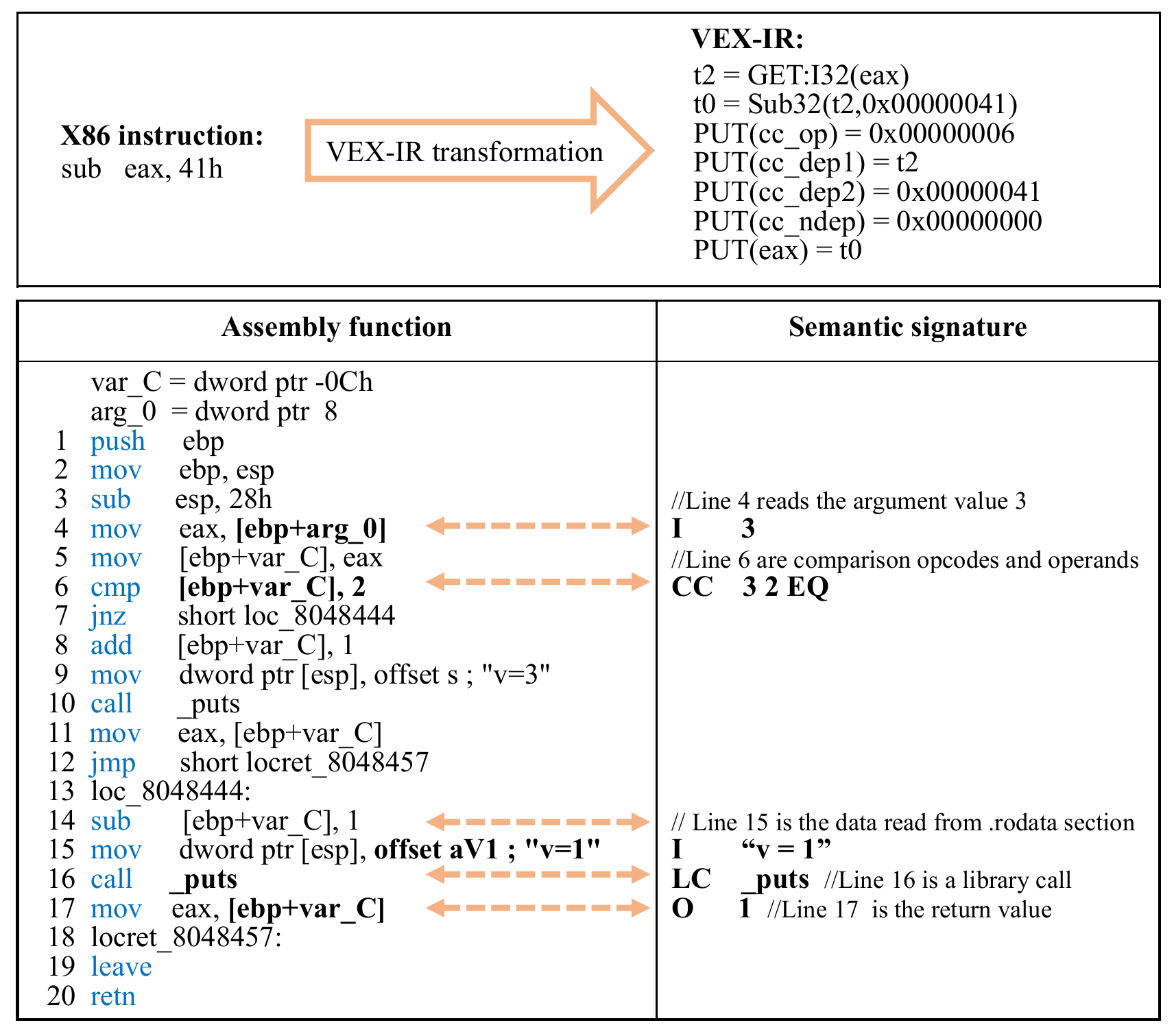}
    \caption{The example assembly function and its semantic signature.}
    \label{fig-signature}
\end{figure}

\textit{BinSeeker} first generates a set of random integers for argument assignment.
For each function, the same random integer sequence is assigned to identified argument registers and stack offsets in turn.
To facilitate the emulation of multiple instruction sets, \tool converts each assembly function into the semantics-preserving \textit{VEX-IR} (intermediate representation) \cite{Valgrind}.
As shown at the top of Fig. \ref{fig-signature}, for a single machine instruction, the conversion process may generate multiple consecutive \textit{VEX-IR} statements.
Unlike a machine instruction that may have multiple consecutive semantic operations, each VEX-IR statement has only one operation and applies to multiple instruction sets.
Through emulating the execution of the function on the \textit{VEX-IR} based on the assigned argument values, we extract unified semantic signatures for the binary functions of different instruction sets.

During the emulation execution, \tool emulates each function individually and records the dynamic execution trace of the function, which we call its semantic signature.
When \tool encounters a call to function $B$ while emulating function $A$, it also enters function $B$ and records the semantic signature of function $B$ in function $A$.
This solves the predictive barrier of function inlining to the semantic learning module.
In addition, \tool excludes the \texttt{main} function to prevent it from encompassing the entire program.
When mutual recursion is encountered, we set a threshold beyond which the function will not be entered again to limit the times for executions of the recursion.
When the same loop is emulated more than the threshold, we reverse the current branch condition to exit the loop and continue the subsequent emulation. The threshold can be dynamically set during emulating, and we refer to \Cacompare \cite{CACOMPARE} to set the threshold for a fair comparison.

The semantic signature consists of four parts: \textit{input values}, \textit{output values}, \textit{comparison opcodes/operands}, and \textit{library function calls}.
The bottom of Fig. \ref{fig-signature} illustrates an assembly function and its semantic signature generated at the end of the emulation.
Here the sample function contains only one stack argument named \textit{arg\_0}, and its corresponding memory location `\textit{[ebp+arg\_0]}' is assigned a value of 3.
\textit{Input values} contain the data read from both the assigned argument values and the data sections (e.g., \textit{.rodata}, \textit{.data}).
The instructions in Lines 4 and 15 of Fig. \ref{fig-signature} contain data reads, and their semantic information is marked as ``\textit{I value}''.
\textit{Output values} consisLt of the return value and memory write values whose addresses are outside the range of the function stack.
Line 17 in Fig. \ref{fig-signature} is the output value of the function when 3 is used as the function argument.
The output values are represented as ``\textit{O value}'' in the semantic signature.
\textit{Comparison opcodes} refer to the condition that controls the jump of basic blocks, and \textit{comparison operands} mean the two values used for comparison.
An example is presented in Line 6 of Fig. \ref{fig-signature}, denoted as ``\textit{CC operands opcode}''.
\textit{Library function calls} record the uses of $C$ language standard library functions during function emulation.
Its semantic information is recorded as ``\textit{LC name}'', such as the Line 16 in Fig. \ref{fig-signature}.

\subsubsection{Similarity Calculation}
After obtaining semantic signatures of the vulnerable function and the top-M candidate functions, \tool uses the Jaccard similarity coefficient to calculate the similarity score as follows: \begin{formula}$J\left ( A,B \right )= \left | A \bigcap B \right | / \left | A \bigcup B \right |$\end{formula}, where $A$ and $B$ are semantic signature sequences of the vulnerable function and the target function.
By descending the similarity scores, \tool reorders the top-M candidate functions and outputs more accurate top-N functions as the final suspected vulnerable functions.

\subsection{Implementation}
For the semantic learning module,
we use \textit{IDAPython} provided by \textit{IDA Pro} \cite{IDA} to create the CFG for each binary function and extract features for each basic block.
Based on the CFG, we infer whether there should be a data dependence edge between two basic blocks by leveraging the \textit{LLVM IR} plugin \cite{miasm} on \textit{IDA Pro}.
We use \textit{TensorFlow} \cite{tensorflow2015-whitepaper} to implement the customized neural network and apply the stochastic gradient descent algorithm to automatically learn model parameters.

For the semantic emulation module,
we use the decompiler of the \textit{IDA Pro} for the X86 and ARM binary.
For the MIPS binary, we use \textit{IDAPython} to traverse CFG paths to obtain register arguments approximately according to the function calling convention.
The key idea is to treat registers that are not written before the first read in the function as register arguments.
We also employ \textit{IDAPython} to traverse assembly instructions of the function to determine the address offsets which are recorded as stack arguments.
Then we utilize \textit{PyVEX} \cite{pyvex} to convert each assembly function into the \textit{VEX-IR} representation \cite{Valgrind}.
\tool emulates each binary function on the \textit{VEX-IR} representation.
After emulating each function contained in the target binary, we record its semantic signature in a separate file.
But for the signatures of vulnerable functions, we store them in a custom data structure into a MongoDB database \cite{MongoDB} to repeatedly perform search tasks.

\section{Experimental Evaluation}
\label{experiment}
\textit{BinSeeker} is based on the similarity of semantic representation (namely, embedding vector in semantic learning and signature in semantic emulation) to complete cross-platform binary vulnerability search.
We compare \tool with three most recent and related state-of-the-art cross-platform binary vulnerability search tools: \Genius \cite{CCS16-Feng} and \gemini \cite{CCS17-Xu} which are semantic learning-based tools, and \Cacompare \cite{CACOMPARE} which is a semantic emulation-based tool.
{\color{black} Furthermore, \toolpre \footnote{\toolpre is similar to our previous implementation \textit{VulSeeker}~\cite{vulseeker}. \textit{VulSeeker-Pro}\cite{vulseeker-pro} is not included in the comparison experiments because it focuses on the optimization of a single architecture.} as the front end learning module of \tool is also involved in the comparison to demonstrate whether we can replace the front end with Genius or Gemini. }
Our experiments aim to answer the following two questions:
\begin{itemize}
    \item RQ1. Is \tool more accurate in predicting similar cross-platform binary functions than the other tools?
    \item RQ2. How efficient is \tool in completing a vulnerability search task?
\end{itemize}

\subsection{Experiment Setup}
\label{sec:experiment_setup}
All experiments are performed on an 8-core 3.60GHz Intel i7 machine with 8G memory, an NVIDIA GeForce 1070 GPU and Ubuntu 14.04 LTS operating system.

\textbf{Data Preparation.}
In order to mitigate experimenter bias and allow for a fair comparison, we prepare the same two datasets as in the other studies \cite{CCS16-Feng,CCS17-Xu,CACOMPARE} as described below to complete different evaluation tasks.
Dataset I is used to train the \toolpre semantic learning model and to directly compare \toolpre with \gemini to explore whether the proposed \toolpre model can achieve better similar code prediction accuracy.
Dataset II is used to evaluate the accuracy and efficiency of \tool for widely studied vulnerabilities.
Since the experiment in the other three tools in comparison involves three architectures, we also include binaries of the same three architectures (\textit{X86}, \textit{ARM} and \textit{MIPS}) for unbiased comparison.
\begin{itemize}
    \item  \textbf{Dataset \uppercase\expandafter{\romannumeral1}}.
    Similar to \cite{CCS17-Xu}, it includes a set of binaries compiled from
    {\color{black}three open-source software}: \textit{BusyBox} (v1.21.0), \textit{OpenSSL} (v1.0.1f and v1.0.1u) and \textit{Coreutils} (v6.5 and v6.7).
    We use two compilers (\textit{GCC} v4.9 and \textit{Clang} v3.4) with four optimization configurations (\textit{O0{\textendash}O3}) to compile these programs to three architectures.
    As a result, we get a total of 368,256 functions and 4,673K basic blocks.
    \item \textbf{Dataset \uppercase\expandafter{\romannumeral2}}.
    We select widely-used real-world programs from those favored for evaluation by other tools in comparison as shown in Table~\ref{tab:benchmark project}, such as \textit{OpenSSL} and \textit{Coreutils} from \Genius \cite{CCS16-Feng}, \textit{curl} and \textit{Wget} from \Cacompare \cite{CACOMPARE}.
    The vulnerabilities in these programs differ greatly in CFGs across platforms and are difficult to detect accurately \cite{CCS16-Feng,CCS17-Xu,CACOMPARE}.
    Each program is compiled into four optimization level versions (\textit{O0{\textendash}O3}) of three architectures using two compilers (\textit{GCC} v4.9 and \textit{Clang} v3.4).
\end{itemize}

\begin{table}[!htbp]
\centering
\scriptsize
\caption{Dataset II: Open-source programs for vulnerability search}
\label{tab:benchmark project}
\begin{tabular}{@{}p{2.1cm}p{1.7cm}p{1.5cm}p{1.5cm}@{}}
\toprule
\textbf{CVE No.} & \textbf{Program}     & \textbf{Module}          & \textbf{Version}        \\ \midrule
2018-11212   & libjepg                  & jmemmgr                  & \multirow{3}{*}{9a}     \\
\cmidrule(r){3-3}
2018-11213   & libjepg                  & \multirow{2}{*}{rdppm}   &                         \\
2018-11214   & libjepg                  &                          &                         \\
\midrule
2018-0494    & \multirow{2}{*}{Wget}    & \multirow{2}{*}{wget}    & \multirow{2}{*}{1.19.1} \\
2017-6508    &                          &                          &                         \\
\midrule
2015-1791    & \multirow{5}{*}{OpenSSL} & \multirow{5}{*}{openssl} & \multirow{5}{*}{1.0.1f} \\
2014-3508    &                          &                          &                         \\
2016-6302    &                          &                          &                         \\
2016-6303    &                          &                          &                         \\
2016-2842    &                          &                          &                         \\
\midrule
2014-9471    & Coreutils                & date                     & 8.13                    \\
\midrule
2017-7407    & curl                     & curl                     & 7.53.1                  \\
\midrule
2015-3237    & \multirow{3}{*}{curl}    & \multirow{3}{*}{libcurl} & \multirow{3}{*}{7.40.0} \\
2015-3145    &                          &                          &                         \\
2015-3144    &                          &                          &                         \\
\bottomrule
\end{tabular}
\end{table}

\textbf{Ground Truth.}
The training of the customized semantic learning model requires a large number of labeled samples of similar and dissimilar binary function pairs.
We use the following strategies to automatically label the samples in the dataset \uppercase\expandafter{\romannumeral1}.
With the source code of the function $f$, we compile it into a set of binary functions denoted as $set(f) = \left \{f_{1},f_{2},\cdots,f_{n}\right \}$ across different implementation platforms.
For each function $f_{i}$ in $set(f)$, we randomly select a different function $f_{j}, i \neq j$ to make up a similar sample, and label them as $\left\langle f_{i},f_{j},+1\right\rangle$.
We also randomly select another binary function $s_{k}$ that is not in $set(f)$ to construct a dissimilar sample, and label them as $\left\langle f_{i}, s_{k},-1\right\rangle$.
A total of 2,761,920 pairs of samples are constructed, where the number of similar sample pairs is half, and no two identical pairs of samples exist.

\textbf{Training Details.}
We apply 10-fold cross-validation to train and evaluate \tool.
Namely, we partition the samples into 10 subsets, each time nine subsets are used to train a model, and one subset is chosen as the test set.
We repeat this 10 times and each time the picked test subset is different. The reported result is averaged over 10 times.
When training the model, 100,000 pairs of samples in nine subsets are randomly selected for training in each epoch.
After finishing each epoch, we randomly shuffle the training set. Note that we refer to the learning component in \tool as \toolm. That is, \toolm produces $M$ candidate functions based on similarity to the function with a known vulnerability.

\textbf{Default Configuration.}
Based on our experiments (discussed in Section \ref{experiments:hyper-parameter}), we set up the hyper-parameters of semantic learning module as follows: the training epoch is 100, the learning rate is 0.0001, the embedding depth $n$ is 2, the embedding size $p$ is 64, the number of iterations $T$ for each basic block is 6 and the size of mini-batch is 10.
For the semantic emulation module, the values of $M$ and $N$ are set to 200 and 25, respectively.

\subsection{Accuracy of Vulnerability Search}
\label{experiments:search accuracy}
Here we mainly answer the RQ1 and focus on whether \tool can identify vulnerabilities across platforms more accurately than other tools.
For each search of the vulnerable function, we obtain the 200 candidate functions from \toolpre and finally choose top-25  functions most likely to have the same vulnerability.

\subsubsection{Overall Result}
We use the vulnerable functions in the \textit{X86-GCC-O3} version of binaries as the source, and the goal of \tool is to identify  functions from the other versions of binaries that have the same vulnerability.
In our experiments, we perform 23 different searches for each vulnerability in dataset II, which results in 345 different searches in total.
Columns 2\textendash6  in Table~\ref{tab:accuracy} show the search ranking of each vulnerability,
and each cell is the average ranking on the 23 different searches, such as X86-Clang-O0 and  MIPS-Clang-O3.

\begin{table}[!htbp]
\scriptsize
\caption{\color{black}The accuracy of five tools for comparison on 15 vulnerabilities.}
\label{tab:accuracy}
\begin{tabular}{@{}lrrrrr@{}}
\toprule
\textbf{CVE No.}                 & \textbf{\textit{Genius}} & \textbf{\textit{Gemini}} & \textbf{\textit{CACompare}} & \textbf{\textit{BinSeeker-}} & \textbf{\textit{BinSeeker}} \\ \midrule
2018-11212                   & 5               & 79              & 1                  & 5                   & 1                  \\
2018-11213                   & 3               & 4               & 1                  & 3                   & 1                  \\
2018-11214                   & 3               & 17              & 1                  & 3                   & 1                  \\
2018-0494                    & 9               & 75              & 3                  & 72                  & 1                  \\
2017-6508                    & 10              & 17              & 68                 & 14                  & 2                  \\
2015-1791                    & 48              & 236             & 67                 & 189                 & 4                  \\
2014-3508                    & 96              & 58              & 60                 & 45                  & 2                  \\
2016-6302                    & 50              & 418             & 19                 & 149                 & 5                  \\
2016-6303                    & 78              & 328             & 236                & 175                 & 12                 \\
2016-2842                    & 149             & 392             & 1                  & 197                 & 1                  \\
2014-9471                    & 1               & 2               & 2                  & 2                   & 2                  \\
2017-7407                    & 5               & 7               & 3                  & 3                   & 2                  \\
2015-3237                    & 23              & 28              & 1                  & 30                  & 1                  \\
2015-3145                    & 8               & 149             & 40                 & 38                  & 5                  \\
2015-3144                    & 69              & 503             & 16                 & 105                 & 1                  \\
\midrule
\textbf{Top-1 (\#@\%)} & \textbf{1@6.67}       & \textbf{0@0}       & \textbf{5@33.33}          & \textbf{0@0}           & \textbf{7@46.67}          \\
\midrule
\textbf{Top-3 (\#@\%)} & \textbf{3@20.00}       & \textbf{1@6.67}       & \textbf{8@53.33}          & \textbf{4@26.67}           & \textbf{11@73.33}          \\
\midrule
\textbf{Top-5 (\#@\%)} & \textbf{5@33.33}       & \textbf{2@13.33}       & \textbf{8@53.33}          & \textbf{5@33.33}           & \textbf{14@93.33}          \\
\midrule
\textbf{Top-20 (\#@\%)} & \textbf{8@53.33}       & \textbf{5@33.33}       & \textbf{10@66.67}          & \textbf{6@40}           & \textbf{100@100}          \\
\midrule
\textbf{MRR} & \textbf{0.17}       & \textbf{0.07}       & \textbf{0.42}          & \textbf{0.13}           & \textbf{0.65}          \\
\bottomrule
\end{tabular}
\end{table}

In Table \ref{tab:accuracy}, we first use the top-$k$ metric to measure the search accuracy of the vulnerable function.
For the 15 vulnerabilities, we count how many times each tool can rank the real vulnerable function in the top-$k$ candidate list and compute their corresponding percentage for each tool in comparison, where the results are recorded in Rows 17\textendash20 and the '@' character is the separator of the number and percentage.
From Table \ref{tab:accuracy}, for different values of $k$, the number of vulnerabilities identified by \tool is significantly more than that of the compared tools.
Specifically, for \Genius, \gemini and \Cacompare, there are only 3, 1 and 8 vulnerabilities ranking in the top-3 candidate list, which results in 20.00\%, 6.67\% and 53.33\% top-3 accuracy, respectively.
On the other hand, \tool identifies 11 real vulnerabilities and achieves a 73.33\% top-3 accuracy rate.
Similarly, the top-5 accuracy for \Genius, \gemini and  \Cacompare are merely 33.33\%, 13.33\% and 53.33\%, respectively, whereas for \tool that is 93.33\%.
All vulnerabilities are ranked in the top-20 candidate list by \tool.
In contrast, the values for the other three tools are 8, 5 and 10, respectively.

In statistics, MRR (mean reciprocal rank) is commonly used to measure the evaluation of the target search results ordered by the probability of correctness, where reciprocal rank value is the multiplicative inverse of the ranking of the first correct result.
In our search scenario, there is only one correct result per search. So we also use this indicator ($MRR = \frac{1}{\left | Q \right |}\sum_{i=1}^{\left | Q \right |}\frac{1}{rank_{i}}, \left | Q \right | = 15$  here) to measure the effectiveness of the four tools.
When each search result is ranked first, the MRR reaches a maximum value of 1.
From Table 3, we can see that \tool has an MRR of 0.65, which is much larger than that of the other four tools.

\subsubsection{Result Analysis}
The ranking in Table~\ref{tab:accuracy} shows that the two learning-based approaches \Genius and \gemini rank four and nine vulnerabilities beyond 50, respectively.
We have examined the assembly functions with the corresponding vulnerability and found the reasons that can cause the inaccuracy.
One is that function inlining exists in the higher optimization level binary functions during compiling, which affects instruction features of the function.
The other reason is that the CFGs of the same function are changeable under different platforms, which is reflected in the semantic embedding vector of the function.
These lead to inaccurate prediction results of these two tools.
Semantic emulation-based approach \Cacompare also ranks 4 vulnerabilities beyond 50.
Because the illegal arguments passed to the functions cause the body of the function to be bypassed during emulation, unrepresentative semantic signatures are produced.
In addition, constant integers in programs are treated as memory references to constant data sections (e.g., \textit{.rodata} section) in one platform, and is directly used as immediate values in another platform.
For binary functions from the same source function, the instruction addressing patterns and the order of instruction accessing memory differ greatly.
This also affects the order and number of semantic signatures.
These cause the phenomenon of mismatching.

\textit{BinSeeker} improves the search accuracy of \gemini by introducing LSFG in the semantic learning component.
Then \tool executes semantic emulation on similar functions to further improve the search accuracy of the vulnerability.
This is why \tool can achieve the best results.
However, one function with the targeted vulnerability is ranking outside the top-10 candidate results in Table \ref{tab:accuracy}.
The reasons that reduce the accuracy of \Cacompare also have effects on \tool, since they all involve the process of function emulation.
Fortunately, \tool only emulates $M$ functions, thus to a certain extent avoiding the shortcomings that affect search accuracy of \Cacompare.
Due to the combination of semantic learning and emulation, \tool can make use of the advantages of both and compensate for the corresponding disadvantages.
As a result, \tool achieves an ideal search ranking in most cases, and a somewhat poor ranking in rare cases.

\subsubsection{Contributions of LSFG and Feature Set}
\label{experiments:LSFG_feature}
We propose a new semantic learning approach in \tool, which has a 1.23$\times$ ranking improvement when compared to the learning-based approach \gemini.
It has a similar process to \gemini, but differs in two aspects.
The first difference is that we also add the DFG between basic blocks instead of just using the CFG.
The other difference is that we use a set of different features to participate in generating function-level embedding vectors.
Thus we conducted experiments on dataset \uppercase\expandafter{\romannumeral1} to verify the contributions of these two improvements to search accuracy improvement.
Based on the default configuration detailed in Section \ref{sec:experiment_setup}, we train our models on the training set according to the following two different requirements, then evaluate the effectiveness on the test set.

\begin{figure}[!ht]
\centering
\begin{minipage}[!htbp]{0.24\textwidth}
     \centering
     \includegraphics[width=1.0\textwidth]{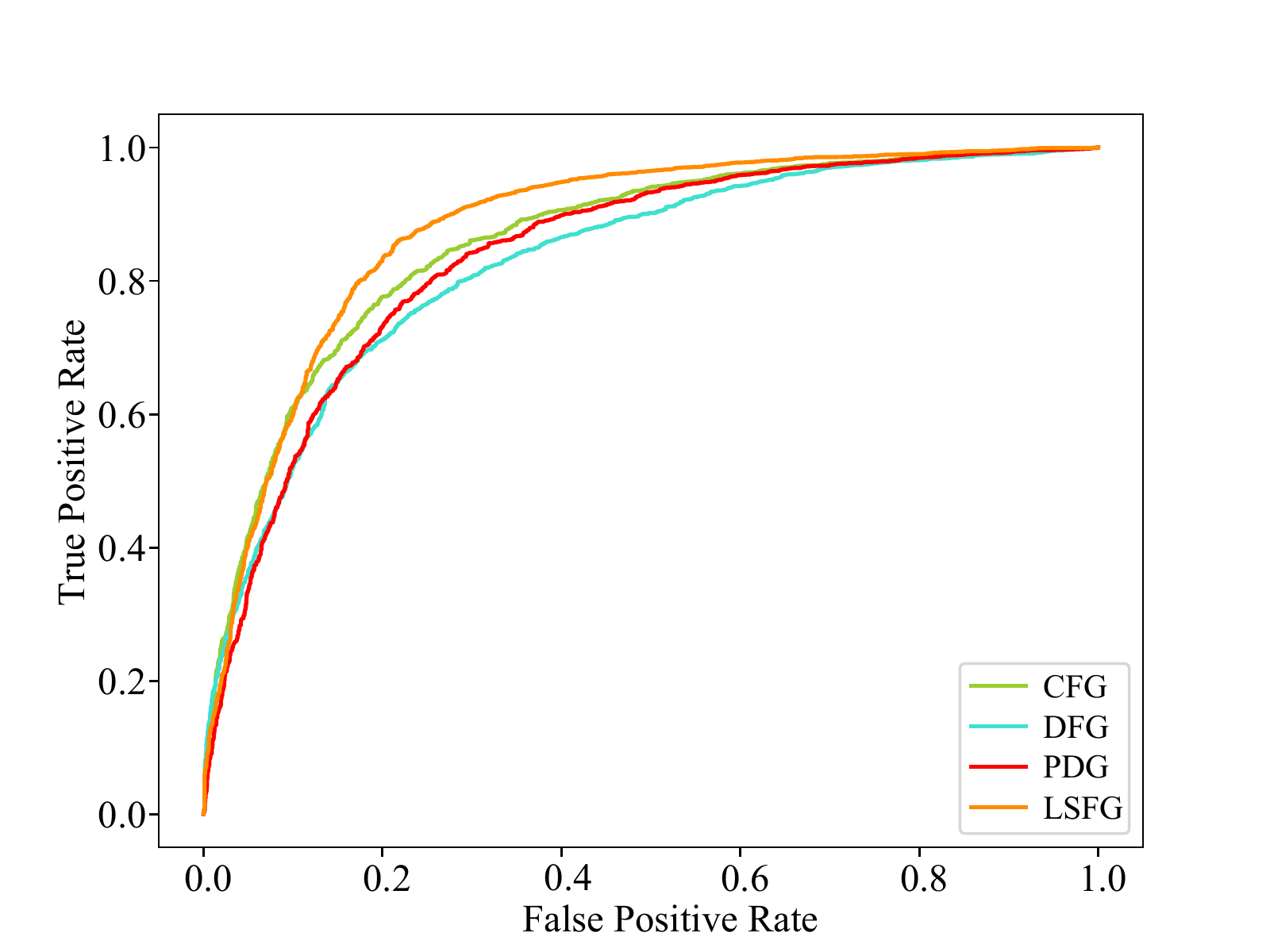}
     \footnotesize{(a) Effectiveness of LSFG}
\end{minipage}
\begin{minipage}[!htbp]{0.24\textwidth}
     \centering
     \includegraphics[width=1.0\textwidth]{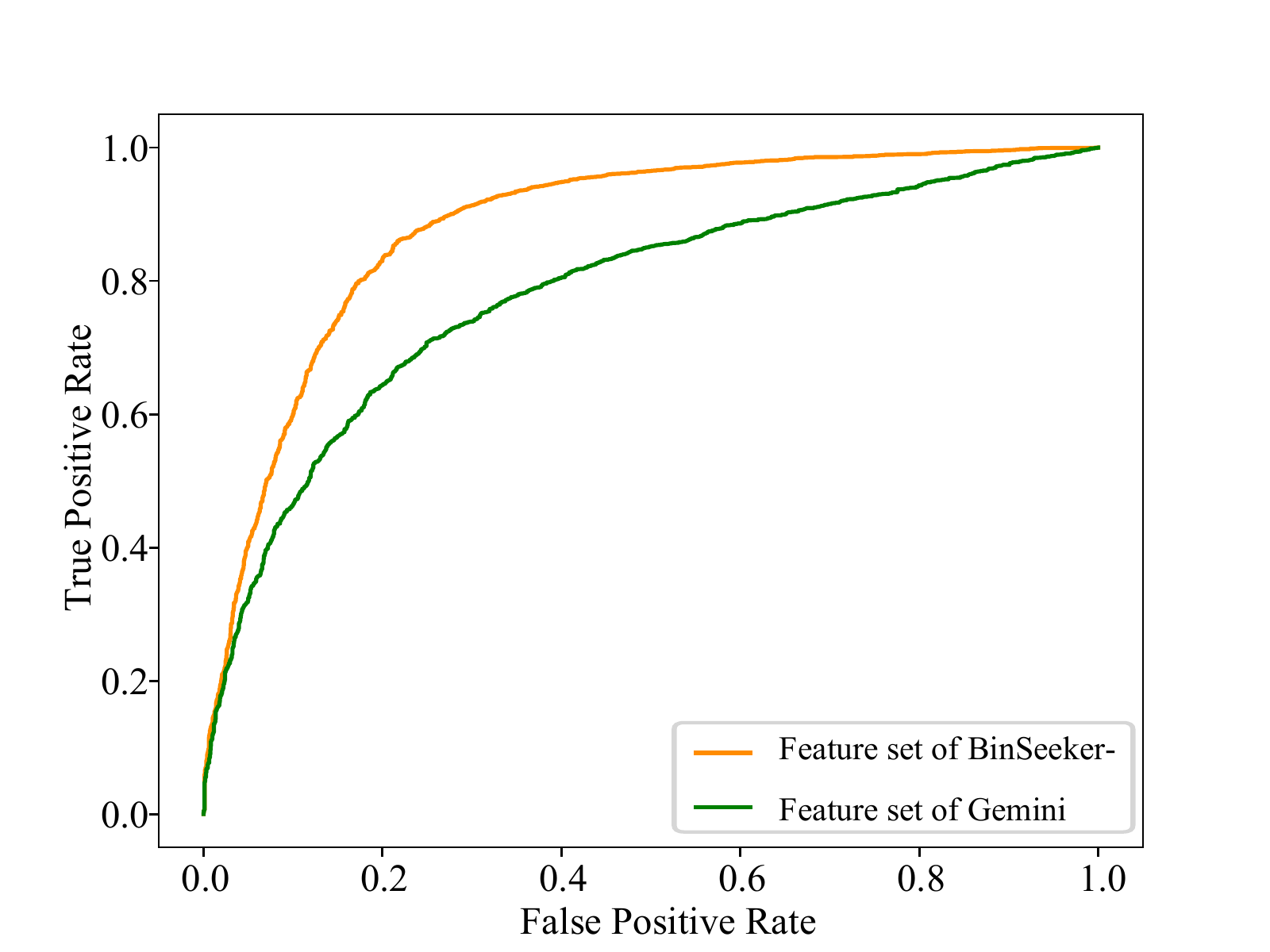}
     \footnotesize{(b) Effectiveness of feature set}
\end{minipage}
\caption{Contributions of LSFG and feature set in \tool}
\label{fig-contributions}
\end{figure}

\textbf{a) Effectiveness of LSFG.}
After fixing the feature set of \toolpre, we use CFG, DFG, PDG and LSFG separately for the experiments.
Fig. \ref{fig-contributions}(a) shows their effectiveness in the form of ROC curve (receiver operating characteristic curve).
The area enclosed by the ROC curve and the $x$-axis is expressed as AUC (area under the curve), which is equivalent to the probability that a randomly chosen positive example is ranked higher than a randomly chosen negative example.
The AUC values for CFG, DFG, PDG and LSFG are 0.86, 0.83, 0.85 and 0.88, respectively.
When the model achieves high classification accuracy, even small improvements are not trivial.
We conclude that applying CFG alone in the semantic learning module can achieve better search accuracy than that of DFG and PDG.
It means that the control flow structure of the function is better at identifying similar functions than data transfer in our semantic learning module.
However, by adding DFG, the search accuracy will be further improved.
This proves that the LSFG used in \tool is effective and achieves a 2.33\% improvement than CFG alone.

\textbf{b) Effectiveness of Feature Set.}
Related researches have proposed several sets of block-level features for predicting similar functions in learning-based methods \cite{CCS16-Feng, alrabaee2016bingold}.
By observing and analyzing the characteristics of different binaries compiled across different platforms, we propose a set of features that are suitable for performing the vulnerability search task.
We use the feature set of \gemini and feature set of \toolpre to conduct experiments for code similarity prediction, both of which are based on LSFG.
Fig. \ref{fig-contributions}(b) shows the performance of different feature sets. 
We observe that the feature set of \toolpre has better performance compared to the feature set of \gemini.
The AUC value for \toolpre's feature set is 0.88, which is 12.8\% higher than the feature set of \gemini.
The experiment shows that the proposed feature set in Section \ref{sec:features} is more robust and change little under various implementation platforms with different microprocessor architectures and various compilation optimization configurations.

\subsection{Time Cost of Vulnerability Search}
\label{sec:efficiency}
We discuss search time obtained by using dataset \uppercase\expandafter{\romannumeral2} and training time which is based on the dataset \uppercase\expandafter{\romannumeral1}, respectively.

\textbf{Search Time. }
We answer RQ2 about how much time \tool needs to complete a vulnerability search task.
This will have a direct bearing on whether \tool can be used in the industry.
We use the \textit{X86-GCC-O0} version of the program as the target to experiment for time cost.
Table~\ref{tab:time efficiency} shows the time cost of the five tools for each program.
Column 2 lists the number of functions in each program.
Columns 3\textendash7 are the time cost to complete a vulnerability search.
All the time is measured in seconds, and the values are rounded.

\begin{table}[!ht]
\centering
\scriptsize
\caption{Time cost of the five tools on each program. (Unit: \textbf{second})}
\label{tab:time efficiency}
\begin{tabular}{@{}L{1.2cm}R{1.0cm}R{0.5cm}R{0.5cm}R{1.0cm}R{0.95cm}R{0.8cm}}
\toprule
\textbf{Program} & \textbf{\#Functions} & \textbf{\textit{Genius}} & \textbf{\textit{Gemini}} & \textbf{\textit{CACompare}} & \textbf{\textit{BinSeeker-}} & \textbf{\textit{BinSeeker}} \\ \midrule
libjpeg          & 580 & 928  & 81 & 343 & 110& 206  \\
Wget             & 804                  & 1,326          & 116           & 531           & 160            & 282           \\
Openssl          & 5,995                 & 8,992          & 849           & 8,432          & 1,145           & 1,323          \\
Coreutils        & 119                  & 198           & 20            & 92            & 23             & 44            \\
curl-7.53.1      & 1,113                 & 1,747          & 167           & 722           & 225            & 362           \\
curl-7.40.0      & 2,760                 & 4,664          & 469           & 1,022          & 549            & 805           \\ \midrule
\textbf{AVG}     & \textbf{1,895}        & \textbf{2,976} & \textbf{284}  & \textbf{1,857} & \textbf{369}   & \textbf{504}  \\ \bottomrule
\end{tabular}
\end{table}

In this experiment, the program contains an average of 1,895 functions.
\gemini, \toolpre and \tool demand less time cost in completing a vulnerability search, which are 284s, 369s and 504s on average, respectively.
It means that they need an average of 0.15s, 0.19s and 0.27s to calculate the similarity between a target function and a vulnerable function.
However, \Genius and \Cacompare require 2,976s and 1,857s to finish one vulnerability search from 1,895 functions.
Their time costs are much higher and require 4.90$\times$ and 2.68$\times$ more than that of \tool.
As a result, \tool is clearly better suited to perform the vulnerability search task on a large scale of code.

Looking closely at Columns 2 through 7 of Table \ref{tab:time efficiency}, we find that the time cost of \Genius, \gemini and \toolpre increases linearly with the number of functions roughly.
The main reason is that these three tools need to extract features, generate semantic embedding vectors, and compute the similarity to the vulnerability for all functions of the program.
However, the time cost of \Cacompare and \tool does not follow the same pattern.
\Cacompare emulates each function dynamically, so the number of loop executions in the function affects the time cost, which does not conform to the linear growth phenomenon.
In contrast to that, \tool only needs to perform the emulation for the fixed number of candidates.
Therefore, the more functions are in the program, the closer the time spent by \tool is to \toolpre.

In summary, for a single function, although the time cost of \tool is 0.12s more than the fastest tool \gemini on average, \tool manages to achieve a better search accuracy in a reasonable amount of time.
Although \Genius is also a learning-based approach, it is nearly 10$\times$ slower than \gemini or \toolpre and thus inappropriate to be the semantic learning module of \tool.

\textbf{Training time. }
Since only the learning-based approaches require training models,
we describe the training time for three tools in comparison: \Genius, \gemini, and \toolpre.
With 100,000 pairs of samples, \Genius requires 50 days to produce a model by spectral clustering.
To train models for 100 epochs under our default configurations, \gemini and \toolpre require 17 and 22 days, respectively.
Nevertheless, the training process is a one-time cost without any effect on vulnerability search efficiency after deployment.
It means that once obtaining an effective model, we can use it in any appropriate scenario without having to retrain it again.
As we will see in Section \ref{experiments:training epochs}, we do not need to train for 100 epochs.
The 50-epoch model can be as good as the 100-epoch model, and the AUC value increases very slowly as the epoch rises so that we can reduce the training time of \toolpre by half (specifically, 11 days).

\begin{figure*}[!htbp]
\centering
\begin{minipage}[!htbp]{0.33\textwidth}
     \centering
     \includegraphics[width=1.0\textwidth]{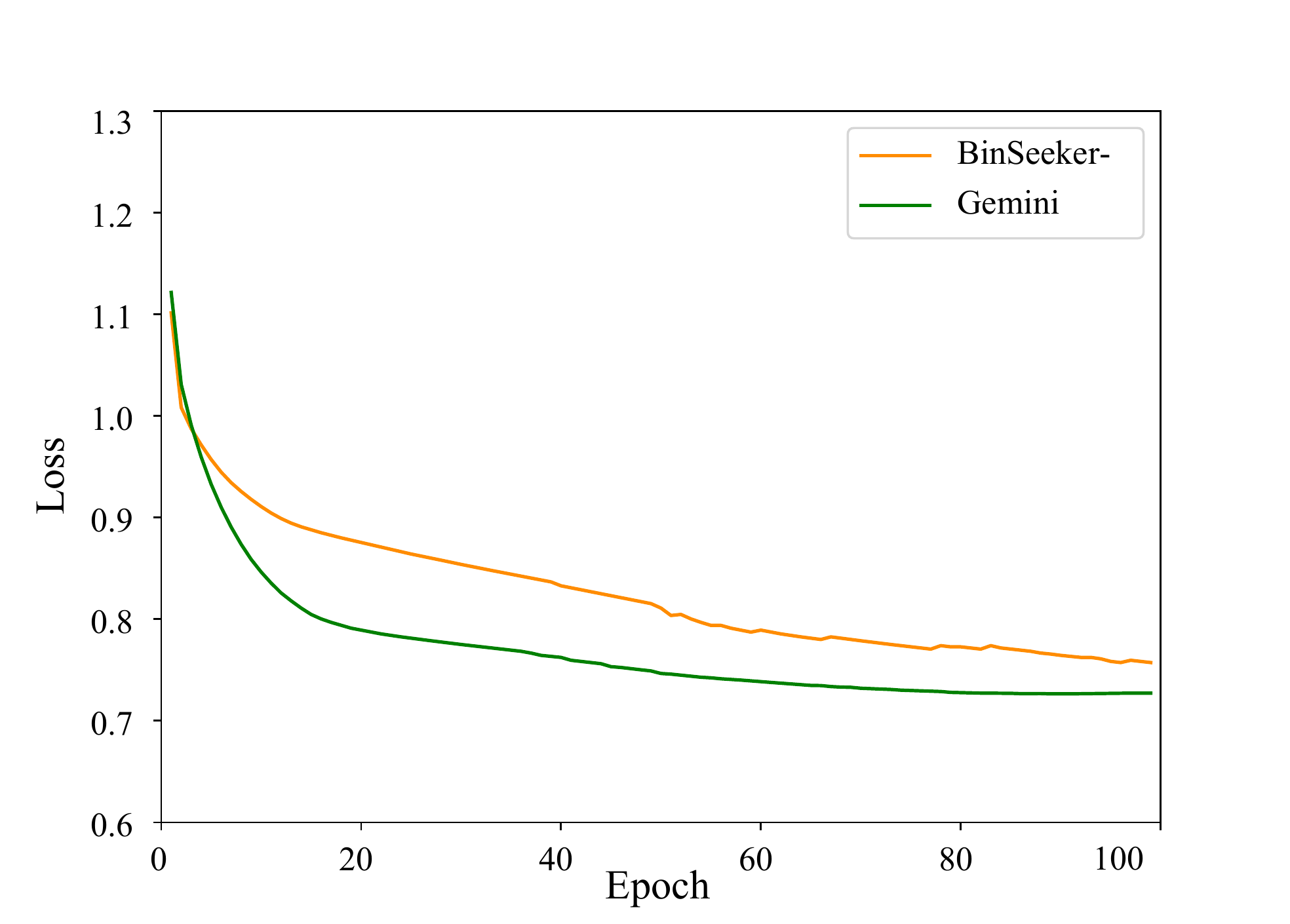}
     \footnotesize{(a) Loss value for different training epochs}
\end{minipage}
\begin{minipage}[!htbp]{0.33\textwidth}
     \centering
     \includegraphics[width=1.0\textwidth]{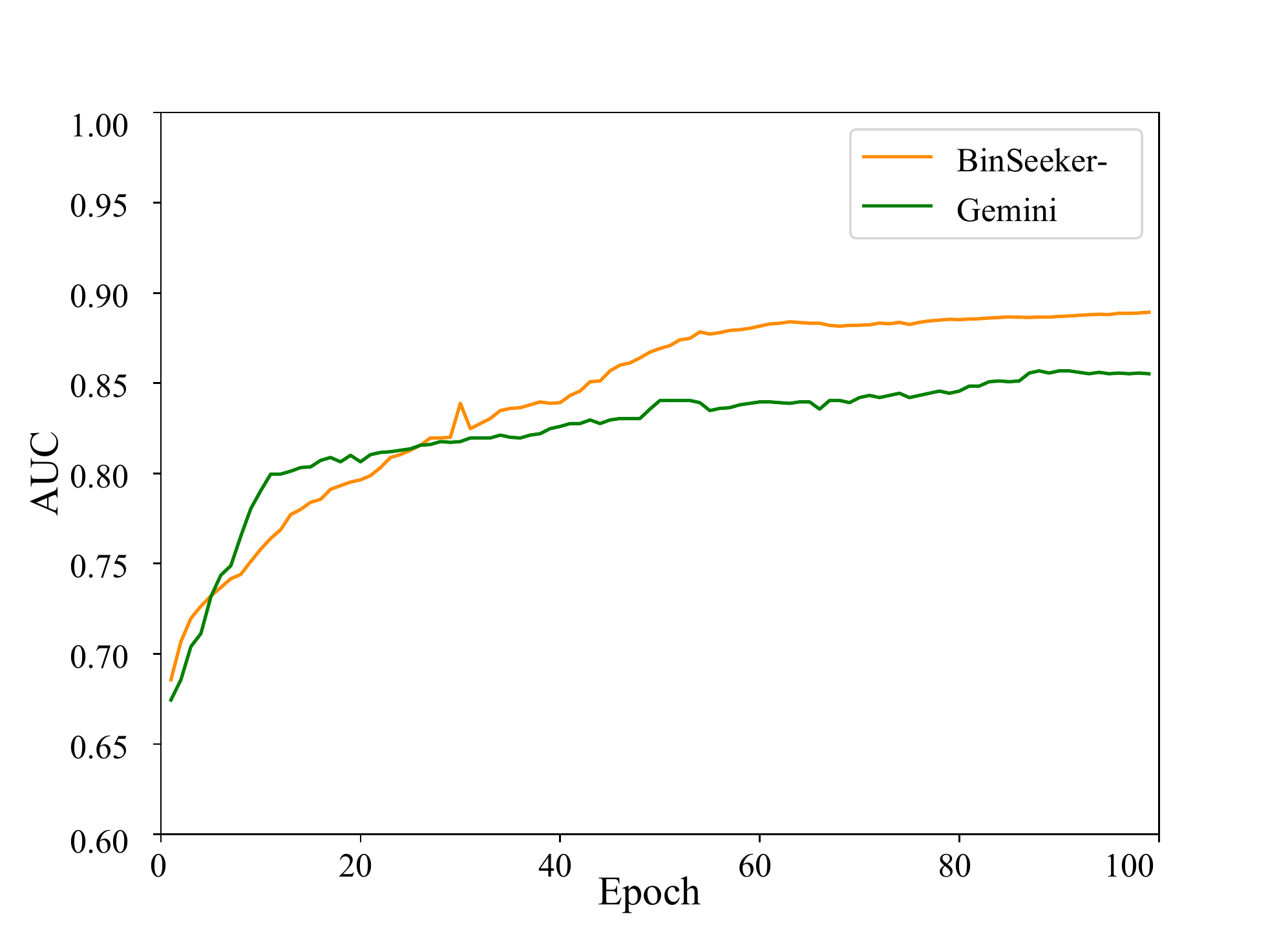}
     \footnotesize{(b) AUC value for different training epochs}
\end{minipage}
\begin{minipage}[!htbp]{0.33\textwidth}
     \centering
     \includegraphics[width=1.0\textwidth]{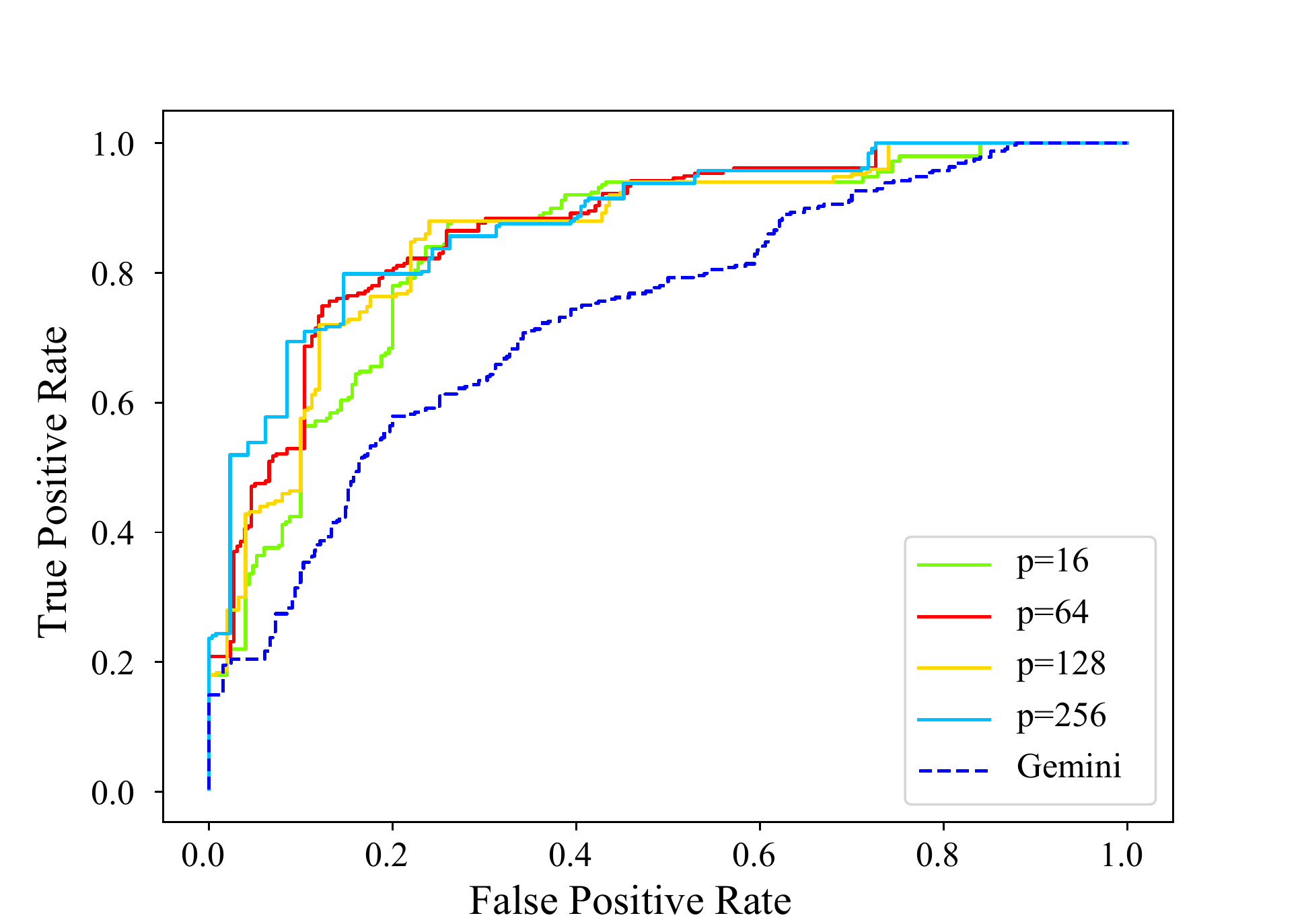}
     \footnotesize{(c) ROC curves for different embedding size p}
\end{minipage}
\begin{minipage}[!htbp]{0.33\textwidth}
     \centering
     \includegraphics[width=1.0\textwidth]{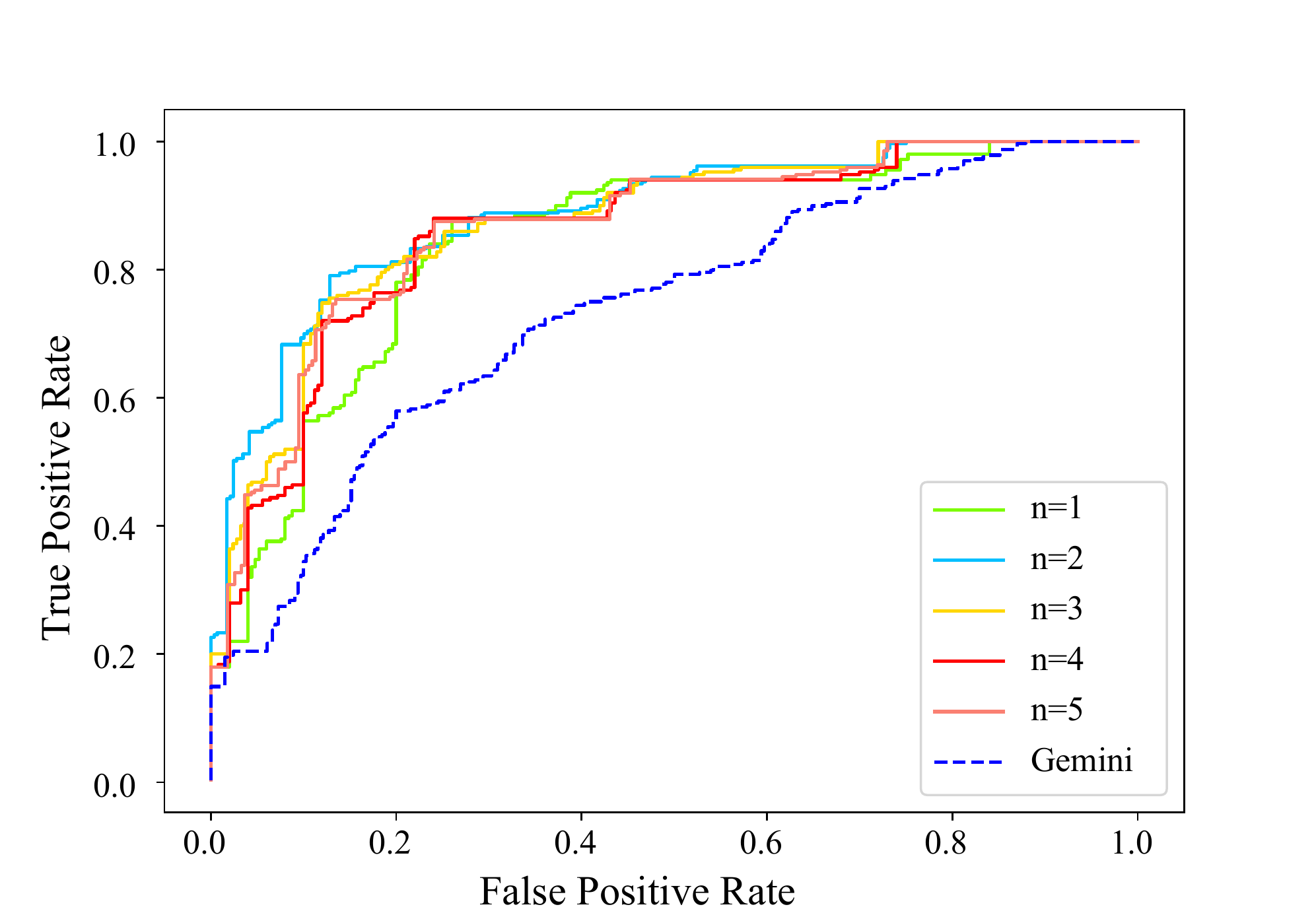}
     \footnotesize{(d) ROC curves for the embedding depth n}
\end{minipage}
\begin{minipage}[!htbp]{0.33\textwidth}
     \centering
     \includegraphics[width=1.0\textwidth]{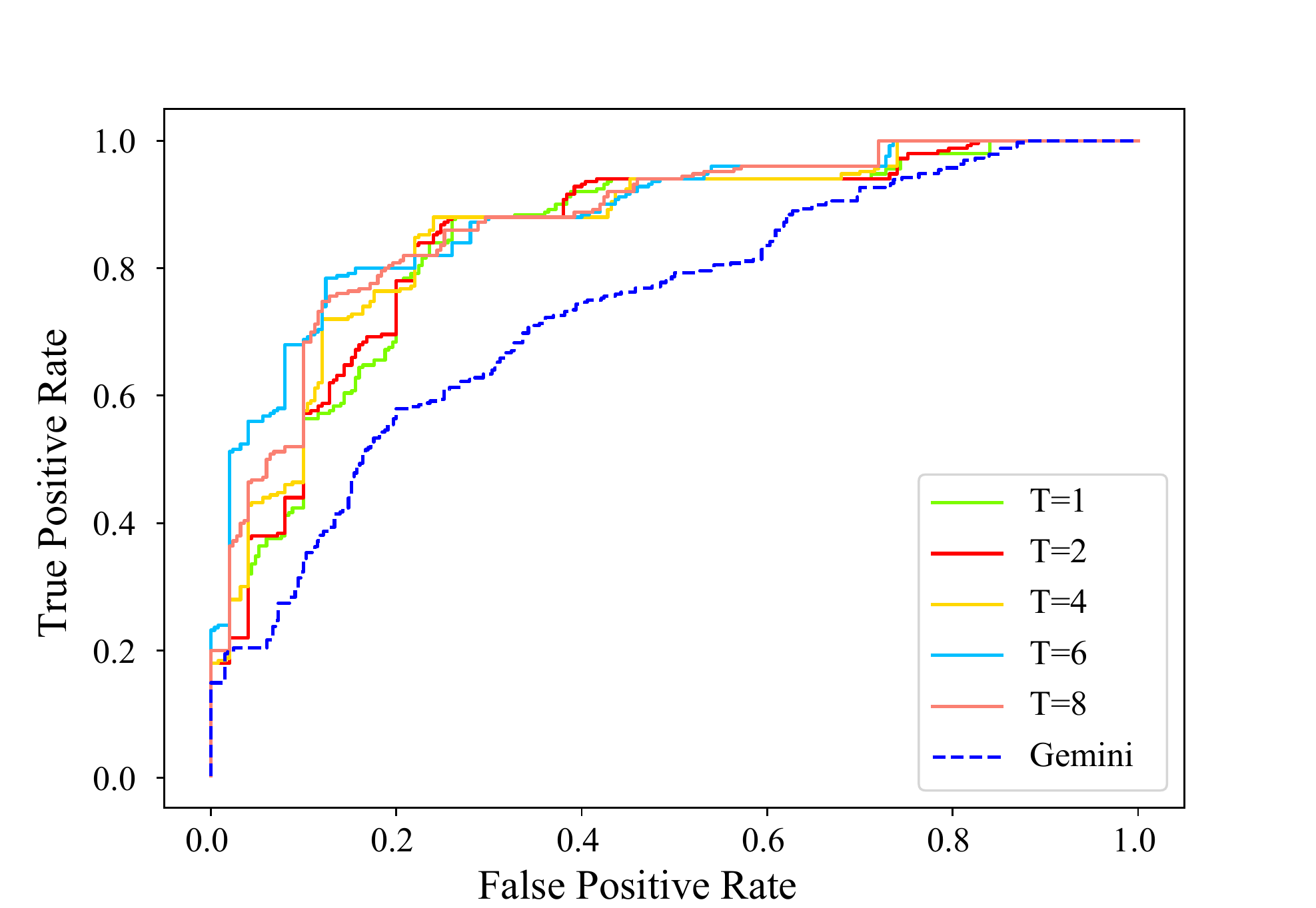}
     \footnotesize{(e) ROC curves for the number of iterations T}
\end{minipage}
\begin{minipage}[!htbp]{0.33\textwidth}
     \centering
     \includegraphics[width=1.0\textwidth]{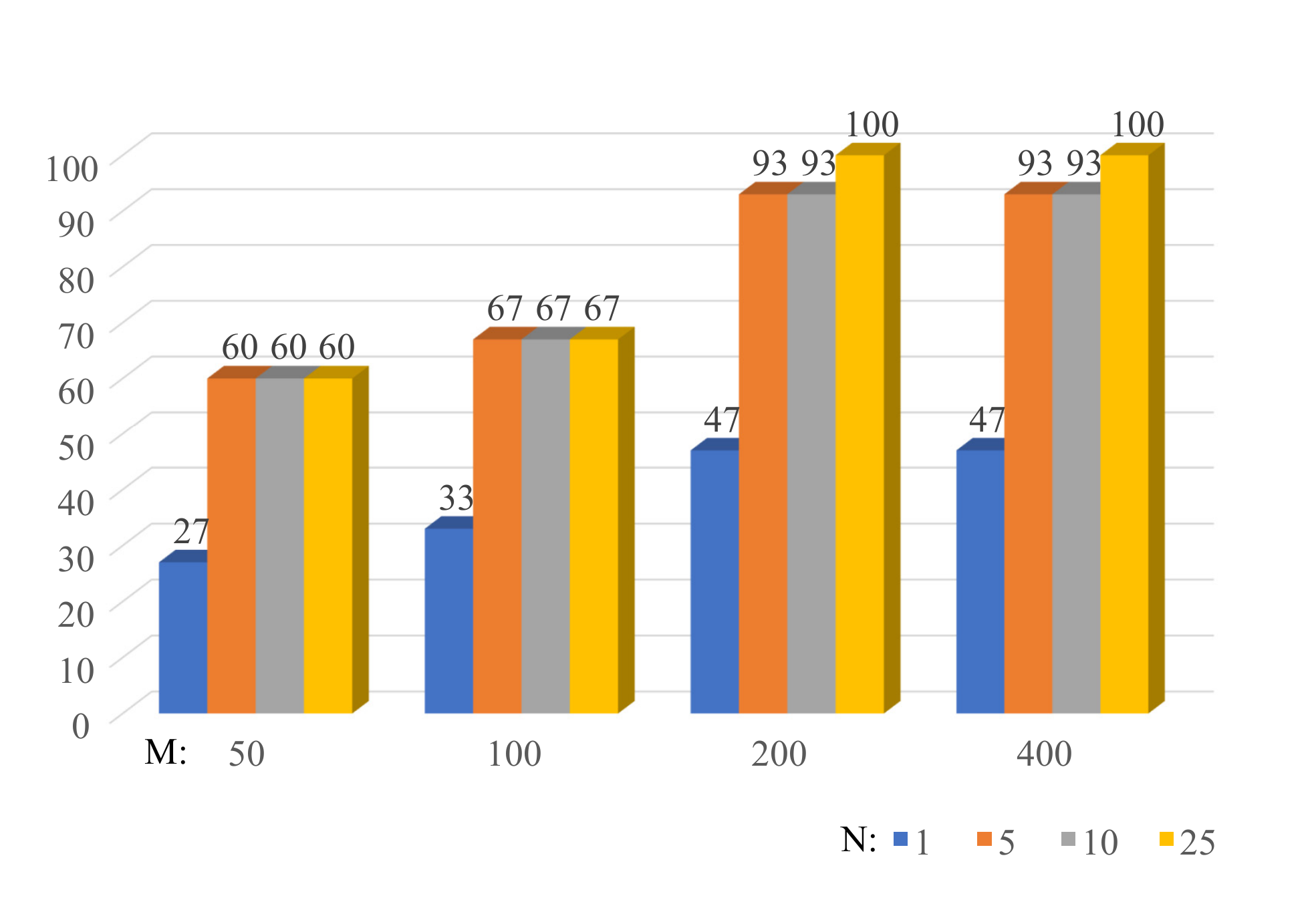}
     \footnotesize{(f) Percentage for different M and N values}
\end{minipage}
\caption{Hyper-parameter studies results.
Employing the test set of dataset \uppercase\expandafter{\romannumeral1},
Fig. \ref{fig-hyperparameters}(a{\textendash}e) describe the predictive effects of the semantic learning model with different choices of the training epochs, the embedding size, the embedding depth and the number of iterations.
Using dataset \uppercase\expandafter{\romannumeral2}, Fig.~\ref{fig-hyperparameters}(f) shows the percentage of vulnerabilities that \tool can successfully find under different $M$ and $N$ values.
With the variation of embedding size, the embedding depth and the number of iterations, the AUC value of \tool remains higher than that of Gemini. The temporary exception in Fig.~\ref{fig-hyperparameters}(b) is because our model contains more parameters and needs more epochs for training in the early stage.
}
\label{fig-hyperparameters}
\end{figure*}

\subsection{Effectiveness of \toolpre Front End }
The search accuracy experiment in Section \ref{experiments:search accuracy} shows that \Genius performs slightly better than \toolpre.
The first intuition is whether we can use \Genius as the front end of \tool to get candidate lists for known vulnerabilities.
Therefore, in this section, we will evaluate this thought from the aspects of accuracy and efficiency by analyzing the experimental data.
The role of the front end is to output a list of candidate functions for further refinement ordering by the back end.
Therefore, changing the value of M will affect whether the output candidate list of the front end can actually contain known vulnerabilities.
The experiments mainly focus on combining \Genius and \toolpre with the semantic emulation back end respectively to evaluate the recall rate and time cost, abbreviated as \textit{Genius + Emulation} vs. \tool.
Since the only difference is the front end, we only change the value of M, and the other configurations are the same as those in Sections \ref{experiments:search accuracy} and \ref{sec:efficiency}.

The effectiveness of the front end in \tool can be evaluated via the recall rate, which measures the ability of the search system to find real vulnerabilities.
In this experiment, the value of $N$ is fixed to 20, $M$ has four values (that is 50, 100, 150, and 200), and the front end output is denoted as top-$M$ candidates.
Let $TP$ be the number of true vulnerabilities ranking in the top-$M$ candidates (true-positives) and $FN$ be the number of true vulnerabilities ranking outside the top-$M$ candidates (false-negatives).
For the false-negative case, \tool would not identify them correctly.
The metric $recall\text{ }rate =\frac{TP}{\left ( TP+FN \right )}$ reflects the completeness of the searched positives on the top-$M$ candidates.
Table \ref{table:recall rate} shows the recall rates with different $M$ values and the average time cost used to calculate each pair of vulnerable function and target function.

\begin{table}[!ht]
\caption{\color{black}The recall rate on different $M$ values and the average time cost.}
\label{table:recall rate}
\begin{tabular}{@{}L{1.8cm}R{0.85cm}R{0.85cm}R{0.85cm}R{0.85cm}R{1.2cm}}
\toprule
\multirow{2}{*}{\textbf{Tools}} & \multicolumn{4}{c}{\textbf{Recall Rate, N=20}} & \multirow{2}{*}{\textbf{Time Cost}} \\ \cmidrule(l){2-5}
                      & \textbf{M=50}    & \textbf{M=100}   & \textbf{M=150}   & \textbf{M=200}   \\ \midrule
\textit{Genius+Emulation}      & 73.33   & 93.33   & 100.00     & 100.00    & 1.57s\\
\textit{BinSeeker}             & 60.00   & 66.67   & 80.00      & 100.00    & 0.19s\\ \bottomrule
\end{tabular}
\end{table}

As can be seen from Table \ref{table:recall rate}, when $N$ is fixed at 20, different $M$ values produce diverse recall rates for the two kinds of combinations between two front ends and one back end.
When $M$ is 50, the combination of \Genius plus emulation can generate a recall rate of 73.33\%, which can identify 22.22\% more real vulnerabilities than \tool.
When $M$ expands to 100, the recall rate of the former reaches 93.33\% and improves 39.99\% than \tool.
However, when $M$ increases to 200, the recall rate of both reaches 100\% in dataset II.
Once the target binary is determined, the total number of binary functions to be processed by each front end is the same and does not change with $M$.
So the average time cost of processing each pair of functions is also an important factor in choosing a suitable front end.
\tool's front end only needs 0.19s to handle a pair of functions, while \Genius takes 7.26$\times$ more time.
This means assuming that a binary has 5,000 functions, it takes \Genius and \toolpre 2.18 hours and 0.26 hours to output top-$M$ candidates to the emulation module, respectively.
In addition, model training time for \Genius is 2.27$\times$ that of \toolpre.

As a summary, when the total number of functions to be searched is determined, the front end that we need to choose should take as little time as possible to gain the maximum recall rate.
Obviously, \toolpre front end with slightly larger $M$ value is more suitable than Genius here.

\subsection{Hyper-parameters Studies}
\label{experiments:hyper-parameter}
In this section, we explore how to set the most appropriate parameters for \tool and evaluate the impacts of different hyper-parameters on search accuracy.
Studies contain two parts: the parameter settings of semantic learning module \toolpre with dataset \uppercase\expandafter{\romannumeral1}, the selection of M and N values in \tool with dataset \uppercase\expandafter{\romannumeral2}.
The first part is from Section \ref{experiments:training epochs} to Section \ref{experiments:iterations},
where all the hyper-parameters are evaluated on the test set.
In the study of each hyper-parameter, except for the parameter being evaluated, other parameters take default settings, as described in Section \ref{sec:experiment_setup}.
The second part is discussed in Section \ref{experiments:value selection}.

\subsubsection{Size of Training Epochs}
\label{experiments:training epochs}
Increasing training epochs results in more time being spent in weight updating.
However, it is pointless to increase the number of training epochs blindly, which may not substantially change the parameter values of the model.
So we want to know when the performance of the model tends to be stable.
In total, we train the model for 100 epochs and evaluate the loss value and AUC value on the test set for every epoch.
The loss value refers to the sum of the squares of the difference between the predicted similarity and the label value among all samples.
Fig. \ref{fig-hyperparameters}(a) and Fig. \ref{fig-hyperparameters}(b) show the loss value and the AUC value, respectively.
We can see that our approach has achieved a good performance in about 50 epochs, the AUC value is 0.88, and the loss value is 0.83.
The loss values for both approaches are below 1.2.
As the number of epochs increases, we eventually have a higher AUC value than Gemini, even though it is lower in about the first 30 epochs.
The main reason is that our model contains more parameters and needs more epochs for training in the early stage.

\subsubsection{Embedding Size}
The embedding size refers to the dimension of the embedding vector used to represent the function semantics.
We use the ROC curves to evaluate which embedding size can achieve the best performance.
Fig. \ref{fig-hyperparameters}(c) plots the experimental results.
When the embedding size is greater than 64, their corresponding ROC curves are close to each other.
We choose 64 as the default embedding size because it can reduce the time cost of training and prediction.
Whatever embedding size is set up for \tool, the AUC value is higher than that of \gemini with the optimal settings.
This phenomenon also applies to the embedding depth and the number of iterations.

\subsubsection{Embedding Depth}
The embedding depth refers to the number of layers of the two fully-connected networks represented as $\sigma _{c}, \sigma _{d}$.
Fig.~\ref{fig-hyperparameters}(d) shows the effects of varying embedding depth.
The relatively good AUC value is obtained when the embedding depth is 2.
This means that by increasing a two-layer fully-connected network, the generated embedding vectors also get higher representation capabilities and better capture the function semantics.
However, when the embedding depth exceeds 2, there will be no benefit other than a higher time cost of training and prediction.

\subsubsection{Number of Iterations}
\label{experiments:iterations}
It refers to the number of hidden layers in the LSFG-based embedding generation network in Fig. \ref{fig-network}(b).
We vary the number of iterations $T$ and get results of ROC curves drawn in Fig. \ref{fig-hyperparameters}(e).
When the number of iterations is 6, our approach achieves the best performance of code clone.
This means that the feature vector of each vertex in LSFG can propagate 6-hops along with the graph topology.

\subsubsection{Selection of $M$ and $N$ Values}
\label{experiments:value selection}
The performance of \tool depends on the $M$ candidate functions produced by the semantic learning module.
This group of experiments will discuss the influence of  $M$ values on the value of $N$  in \tool.
Fig. \ref{fig-hyperparameters}(f) shows the effectiveness of different $M$ and $N$ values.
The $x$-axis is the different $M$ values (50, 100, 200, and 400).
The $y$-axis is the percentage of vulnerabilities that \tool can successfully find when the M and N values are fixed.
For each M, we study the percentage of four N values (1, 5, 10, and 25).

From Fig. \ref{fig-hyperparameters}(f), we know that when N is fixed, the number of vulnerabilities \tool can detect increases with the increase of the M value.
But the N value needs to be at least 25 to achieve 100\% search accuracy.
Larger M value is more likely to ensure that the vulnerability can be searched, which is cost-effective to increase the time cost of just 0.98s per function on average.
When the M value is fixed, the percentage also gets higher with the increase of the N value.
When M is 200, the percentage can reach 100\%.
However, it may not grow to 100\% due to a small M value that will result in no vulnerability in the top-M candidate results.
To sum up, the values of M and N are best set to 200 and 25 for all vulnerabilities to be found within a relatively short time.

\section{Threats to Validity}
\label{threat-to-validity}
We present some potential threats that may affect the performance of \tool and provide some coping strategies.

\textbf{a) Semantic Learning Model.}
The performance of \tool depends on the top-M candidate functions exported by the semantic learning module.
If we can reduce the value of M without losing accuracy, the total vulnerability search time required by \tool will be shortened.
When training the network model, we can enhance the generalization ability of the model by increasing the discrete training samples from multiple binary programs.
In general, the model with a larger training epoch will have stronger vulnerability prediction ability.
Nevertheless, we need to pay attention to  model over-fitting.

\textbf{b) Function Inlining.}
{\color{black}Compiler optimizations may inline some functions to the callers to maximize the runtime speed.}
Whether the analyzed function contains inlined functions affects the number of instructions within the function, and ultimately affects function embedding vectors.
Bingo \cite{Bingo} proposes an inline decision algorithm based on six commonly-observed invocation patterns.
It is possible to adopt the algorithm to inline called functions selectively, then to extract basic block features.
In this way, analyzed cross-platform functions will have a consistent inline strategy, which is likely to increase search accuracy.

\textbf{c) Binary Obfuscation.}
Code obfuscation can significantly change the control flow structure (e.g., CFG flattening, opaque predicates), sometimes even the data dependence structure,
which brings enormous side effects to the function semantics.
Before constructing the LSFG for learning and emulating, we first need to deobfuscate code \cite{Deobfuscation}, which will effectively solve the impact of obfuscation.

\textbf{d) Binary Diversity.}
The binary functions obtained from the same source code under different compilation scenarios have great differences.
For some less complex compilation scenarios, the accuracy of \gemini and \tool can be much higher than the results presented in Fig. \ref{fig-hyperparameters}(b). For example, in the dataset settings of \cite{CCS17-Xu}, \gemini's AUC value of the model is 0.971 and ours is about 0.984, which is much higher than the value of 0.818 and 0.885 tested in our test set.
Our dataset \uppercase\expandafter{\romannumeral1} contains five different programs of various sizes with more compilation options, but the dataset of \gemini only contains two.
We find that the more complex the compilation scenario is, the greater the improvement \tool achieves.
The emulation approach described in this paper mitigates the impact of binary diversity to some extent.

\textbf{e) False Negative.}
 If the front end \toolpre gets some false negative, the second part is useless. But both learning-based and emulation-based approaches have the problem of false negatives and false positives. The false positives of learning-based approaches are more serious, and the false negatives of emulation-based approaches are more serious. To our best knowledge, it is not possible to completely solve the problem of false positives and false negatives currently, and there is no exhaustive solution for every single one. We can increase the parameter M to reduce the possibility of false negative cases, as described in Section 4.5.5. In the case of the original Gemini, if we consider a function with vulnerability ranked top-5 as accurate, Gemini only achieves an accuracy rate of 13.33\%.  When the values of M and N are 100 and 5, the false negatives of BinSeeker is only 33\%, and the top-5 accuracy is 67\%. From the figure, we know that when N is fixed, the number of vulnerabilities BinSeeker can detect increases with the increase of the M value. Larger M value is more likely to ensure that the vulnerability can be in the candidate list, which is cost-effective to increase the time cost of just 0.98s per function on average. In our experiments, when the values of M and N are set to 200 and 25, all vulnerabilities can be found in the top-25 results. This is a significant breakthrough in reducing the human efforts of manually confirming real vulnerabilities from a large number of suspicious functions.

\section{Related Work}
\label{related-work}
\subsection{Syntax and Structure Based Search}
The idea of vulnerability search based on syntax and structure is that similar code fragments have similar syntax structures.
\textit{CCFinder} \cite{Kamiya2002CCFinderAM} detects cloned source code in large scales based on lexical code tokens.
It generates the token sequences of the source code through a lexical analyzer and then obtains the regularized sequences with rule-based transformation.
Finally, it applies a suffix-tree matching algorithm to compute similarity.
\textit{BinDiff} \cite{Flake2004StructuralCO} builds CFGs of the two binaries and then adopts a heuristic algorithm to normalize and match the two CFGs.
\textit{BinSlayer} \cite{Bourquin2013BinSlayerAC} improves \textit{BinDiff} by adopting the \textit{Hungarian algorithm} \cite{hungarian_algorithms} for bipartite graph matching.
\textit{DECKARD}\cite{Jiang2007DECKARDSA} produces an abstract syntax tree (AST) to represent the source program, and further extracts feature vectors from AST, improving the efficiency and accuracy of the detection.

\subsection{Semantics Calculation Based Search}
Semantics calculation based vulnerability search approaches use the semantic features calculated from the syntactic structures to better represent searched codes.
\textit{COP} \cite{Luo2014SemanticsbasedOB} is a plagiarism
detection tool that combines program semantics with the longest common sub-sequence based fuzzy matching.
\textit{BinHunt} \cite{Gao2008BinHuntAF} considers matching CFGs as the maximum common induced sub-graph isomorphic problem.
It leverages symbolic execution and theorem proving to match the basic blocks with the same semantics.
\textit{BLEX} \cite{Egele2014BlanketED} is a dynamic function matching tool that uses several semantic features obtained during the function execution (for example, values read from the program heap) in the matching process.
\textit{BinGold} \cite{alrabaee2016bingold} extracts the semantics of binary code concerning both data and control flow and synthesizes them into a novel representation called the semantic flow graph.
However, it does not support cross-architecture clone detection, and its average precision is 74.97\%.
\textit{BinSim} \cite{Ming2017BinSimTS}  calculates the equivalences of aligned system calls to better handle code obfuscation. It is a hybrid method to identify fine-grained semantic similarities or differences between two execution traces.

\subsection{Learning Based Search}
Learning-based vulnerability search approaches automatically learn semantic features of the program or select proper code  similarity algorithms to find the cloned code.
\textit{VulPecker} \cite{Li2016VulPeckerAA} applies the SVM classification approach to select a set of code-similarity algorithms that could distinguish unpatched pieces of code from the patched ones.
\textit{VulDeePecker} \cite{Li2018VulDeePeckerAD} uses code gadgets to represent programs and employs BLSTM neural network to extract features instead of having human experts manually defining the features.
In \cite{White2016DeepLC}, the authors present a deep learning-based clone detection tool, extracting hidden patterns of the lexical and syntactic levels based on the RNN.
Based on the bipartite graph matching algorithm, \textit{Genius} \cite{CCS16-Feng} calculates the similarity between a specified ACFG (attributed control flow graph) and each representative ACFG in the codebook generated by the spectral clustering algorithm.
\textit{Gemini} \cite{CCS17-Xu} generates an embedding vector for each function represented by the CFG, then compares each pair of vectors to get the prediction result.
\textit{VulSeeker} \cite{vulseeker} proposes a set of lightweight instruction features and integrates DFG into CFG to enhance the robustness against structural differences in the CFG.
\textit{VulSeeker-Pro} \cite{vulseeker-pro} supplements multiple semantic signatures to evaluate function similarity, and mainly focuses on the optimization of a single architecture.
\tool seamlessly integrates our previous optimization  \cite{vulseeker,vulseeker-pro}, and supports the cross-architecture function emulation to achieve a better performance in both time and accuracy.

\section{Conclusion}
\label{conclusion}
In this paper, we present \tool, an accurate and efficient cross-platform binary vulnerability seeker that integrates semantic emulation with semantic learning.
In semantic learning, by combining both the data flow dependency and the control flow dependency of the binary function, we capture more function semantics than the existing approaches and output the $M$ candidate functions that are similar to the vulnerable function.
Then through semantic emulation, \tool further improves the search accuracy and outputs more accurate top-$N$ candidate functions out of the $M$ candidates.
Overall, \tool achieves higher search accuracy with a lower computation requirement than state-of-the-art tools such as \Genius, \gemini and \Cacompare.
Compared to the time users spend in manually identifying real vulnerabilities from a collection of hundreds of false positives and a few true positives, the running time is almost negligible. Our future work will seek to improve the robustness of the semantic emulator and apply it to more platforms.

\section*{ACKNOWLEDGMENTS}
This work was sponsored in part by the NSFC Program under Grant 61527812, and in part by the National Science and Technology Major Project of China under Grant 2016ZX01038101.




%

\vspace{-1cm}
\begin{IEEEbiography}[{\includegraphics[width=1in,height=1.25in,clip,keepaspectratio]{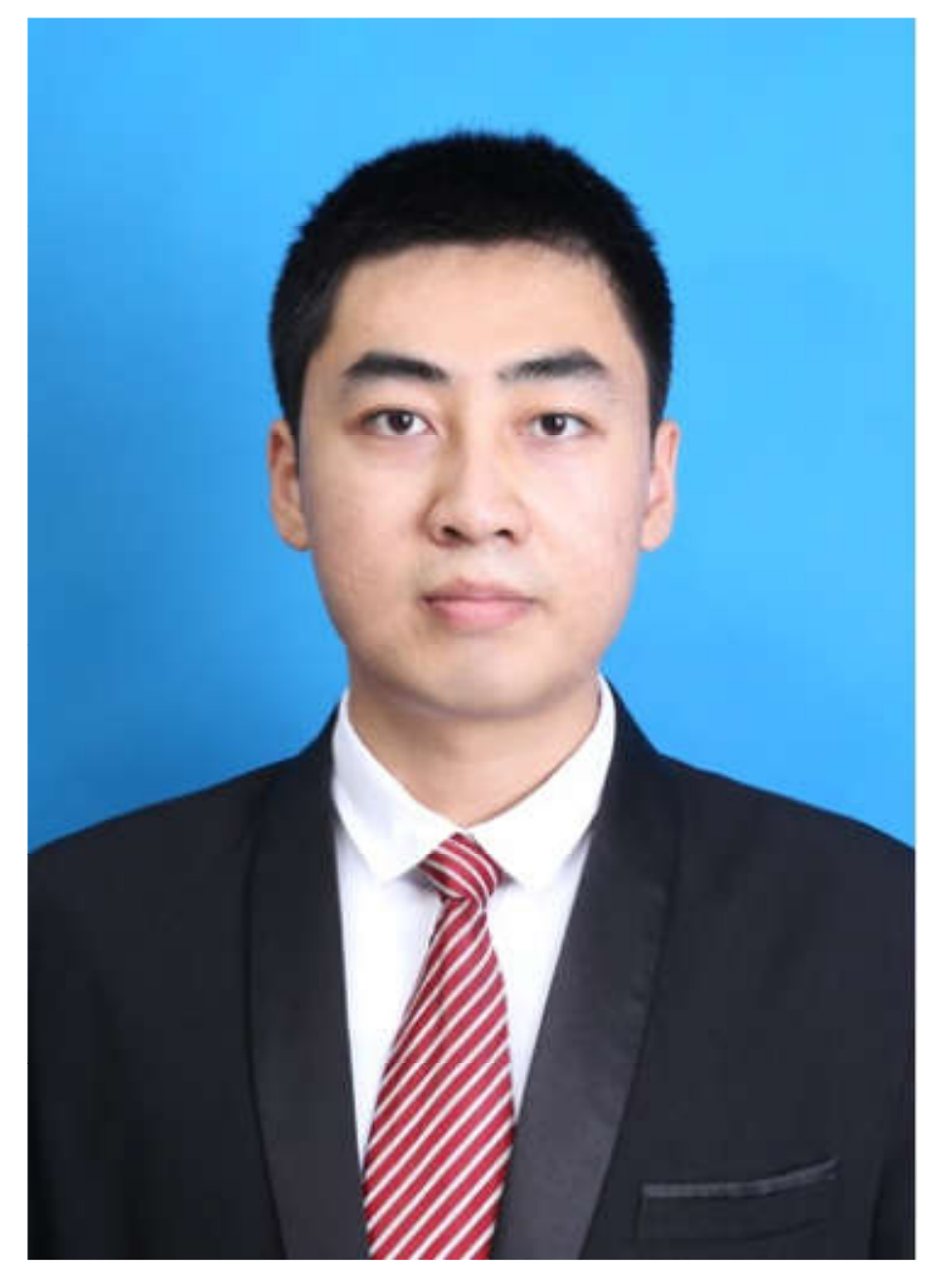}}]{Jian Gao}
received the BS degree in software engineering from Beijing University of Posts and Telecommunications, Beijing,
China, in 2016.
He is currently working toward the Ph.D. degree in software engineering at Tsinghua University, Beijing, China.

His research interests include binary vulnerability search, binary clone detection, machine learning in program analysis and their applications to industry.
\end{IEEEbiography}

\vspace{-0.5cm}
\begin{IEEEbiography}[{\includegraphics[width=1in,height=1.25in,clip,keepaspectratio]{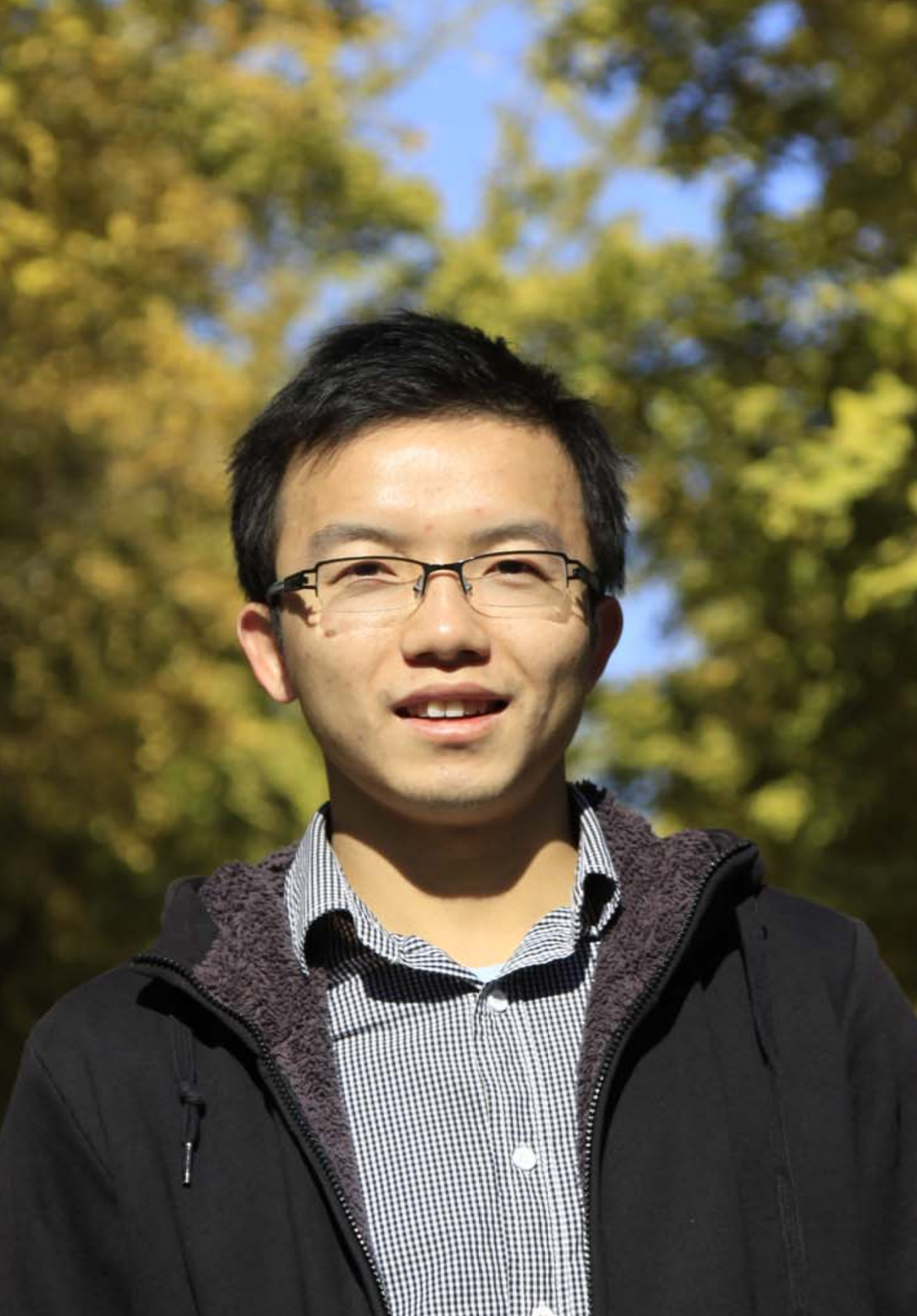}}]{Yu Jiang}
received the BS degree in software engineering from Beijing University of Posts and Telecommunications in 2010, and the PhD degree in computer science from Tsinghua University in 2015.
He was a Postdoc researcher with the Department of Computer Science, University of Illinois at Urbana-Champaign, Champaign, IL, USA, in 2016, and is now an assistant professor in Tsinghua University.
His current research interests include domain specific modeling, formal computation model, formal verification and their applications in embedded systems.
\end{IEEEbiography}

\vspace{-0.5cm}
\begin{IEEEbiography}[{\includegraphics[width=1in,height=1.25in,clip,keepaspectratio]{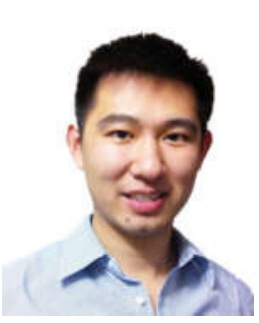}}]{Zhe Liu}
received his Ph.D degree from the Laboratory of Algorithmics, Cryptology and Security (LACS), University of Luxembourg. His Ph.D thesis has received the prestigious FNR Awards. He is a full professor in College of Computer Science and Technology, Nanjing University of Aeronautics and Astronautics (NUAA) and SnT, University of Luxembourg. He has been a visiting scholar in City University of Hong Kong, COSIC, K. U. Leuven as well as Microsoft Research, Redmond. His research interests include computer arithmetic and information security.
\end{IEEEbiography}

\begin{IEEEbiography}[{\includegraphics[width=1in,height=1.25in,clip,keepaspectratio]{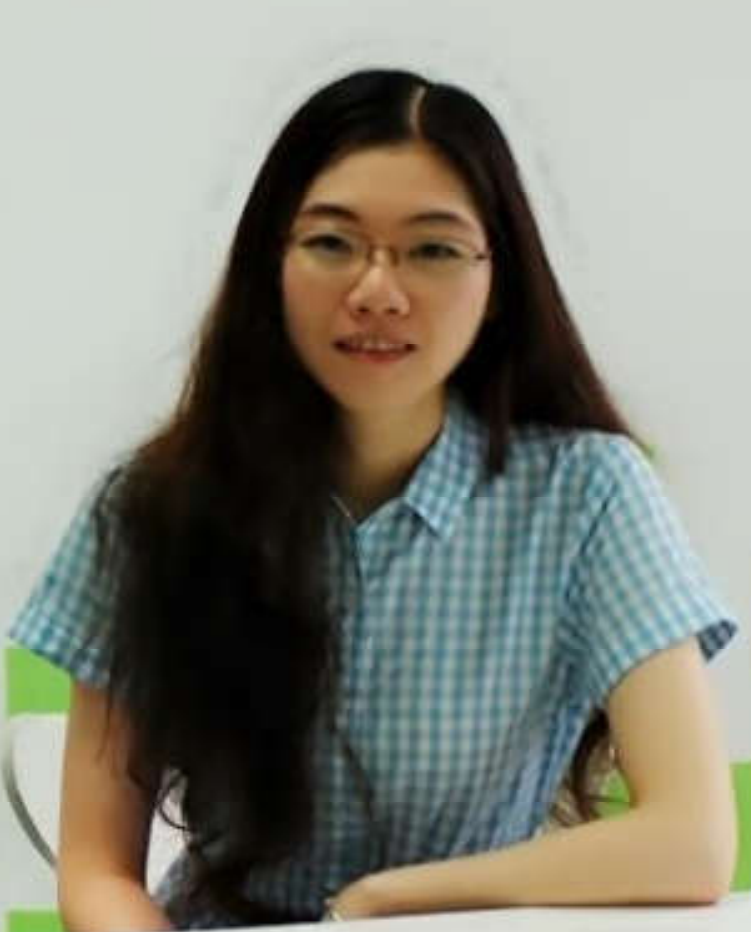}}]{Xin Yang}
received the BS degree in software engineering from Beijing Jiaotong University, Beijing,
China, in 2016.
She is currently working toward the M.S. degree in software engineering at Tsinghua University, Beijing, China.

Her research interests include binary vulnerability search, machine learning in program analysis and their applications to industry.
\end{IEEEbiography}

\begin{IEEEbiography}[{\includegraphics[width=1in,height=1.25in,clip,keepaspectratio]{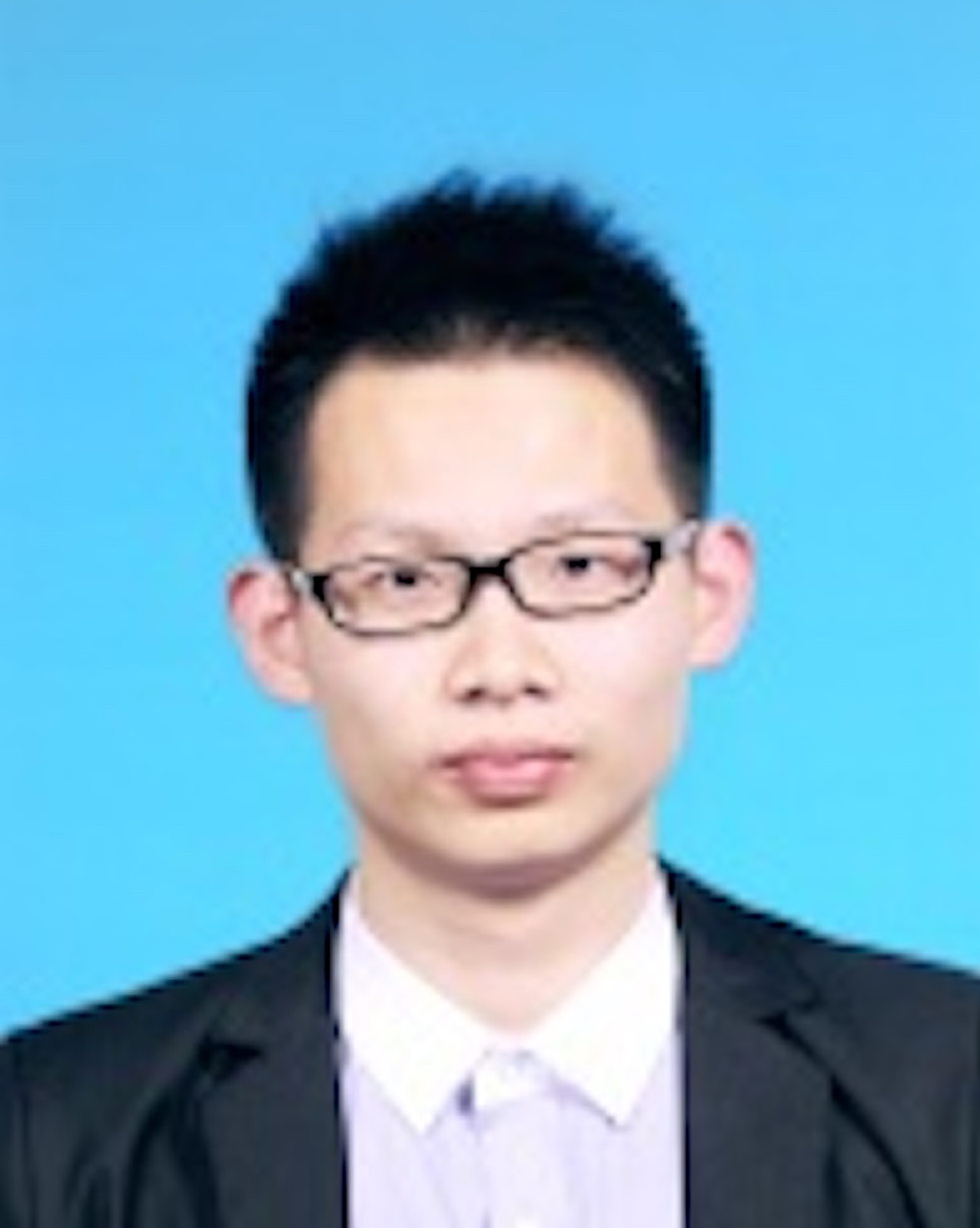}}]{Cong Wang}
received his BS degree in the School of Software, Tsinghua University, Beijing, China, in 2015.
He is currently working towards the Ph.D. degree in software engineering at Tsinghua University, Beijing, China.

His research interests include program analysis, software testing and deep learning.
\end{IEEEbiography}

\vspace{-4cm}

\begin{IEEEbiography}[{\includegraphics[width=1in,height=1.25in,clip,keepaspectratio]{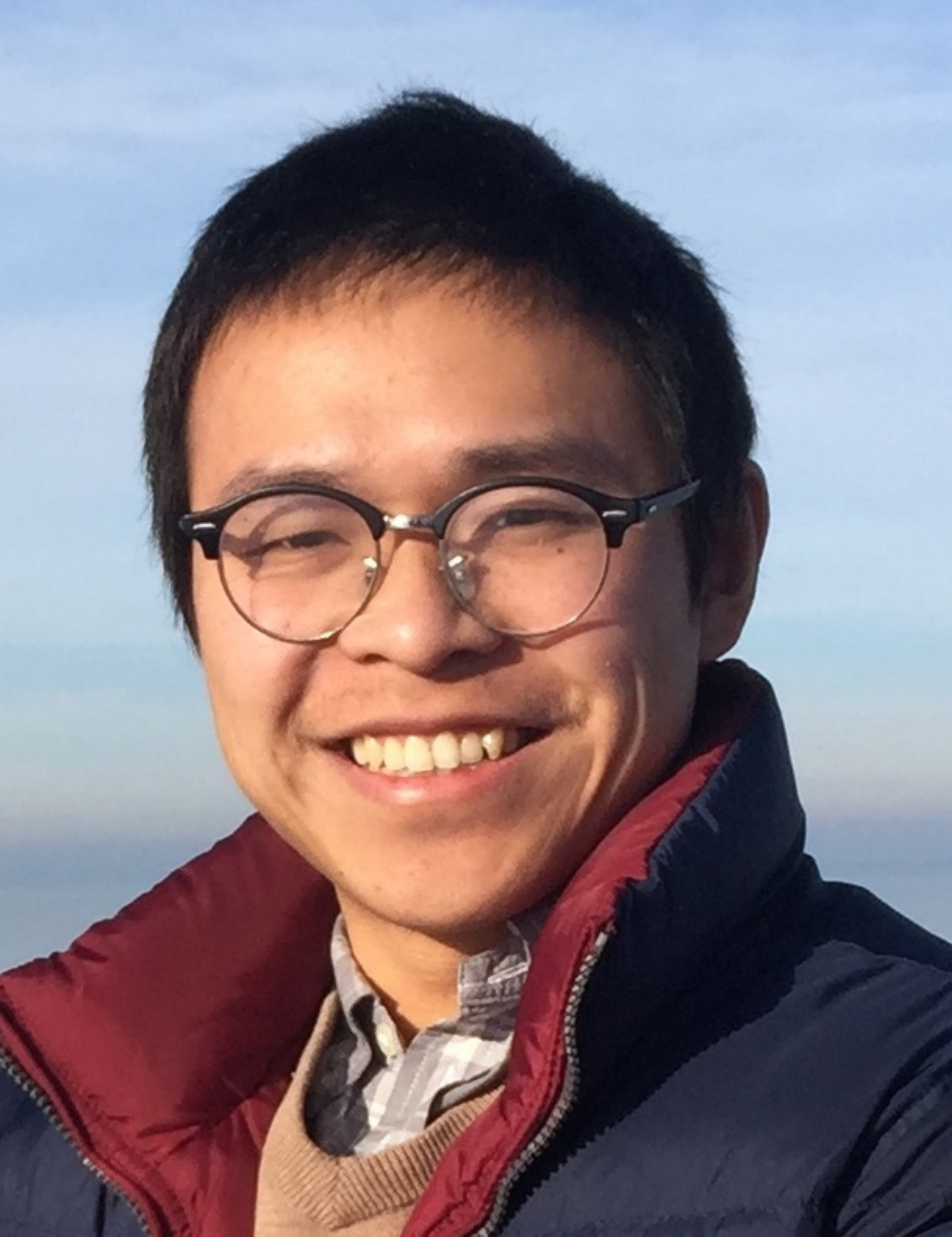}}]{Xun Jiao}
is an assistant professor in the ECE department of Villanova University.
He obtained the Ph.D. degree from the department of Computer Science and Engineering at the University of California, San Diego. He received the dual bachelor's degree from the Beijing University of Posts and Telecommunications, China and the Queen Mary University of London, United Kingdom, in 2013. His research interests include error-tolerant computing and machine learning.
\end{IEEEbiography}

\vspace{-4cm}
\begin{IEEEbiography}[{\includegraphics[width=1in,height=1.25in,clip,keepaspectratio]{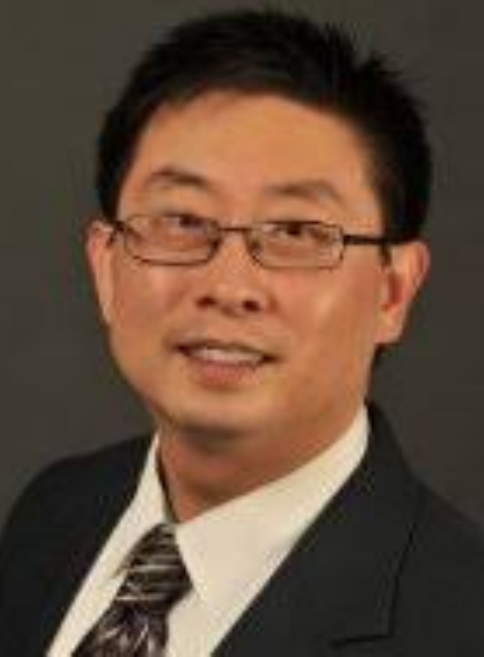}}]{Zijiang Yang}
is a professor in the CS department at Western Michigan University. He holds a Ph.D. from the University of Pennsylvania, an M.S. from Rice University and a B.S. from the University of Science and Technology of China. Before joining WMU he was an associate research staff member at NEC Labs America.
His research interests are in the area of software engineering with the primary focus on the testing, debugging and verification of software systems. He is a senior member of IEEE.
\end{IEEEbiography}

\vspace{-4cm}
\begin{IEEEbiography}[{\includegraphics[width=1in,height=1.25in,clip,keepaspectratio]{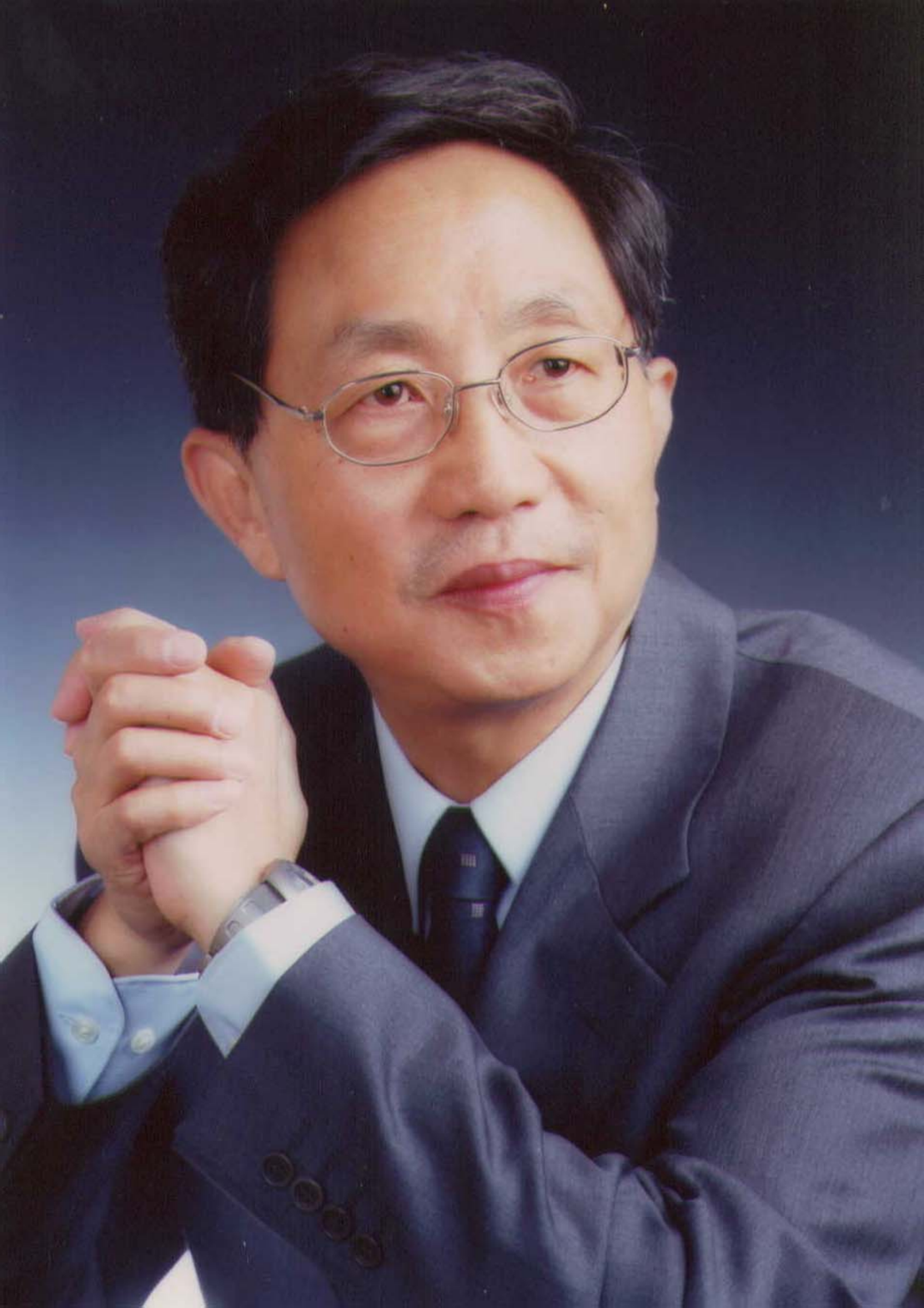}}]{Jiaguang Sun}
received the BS degree in automation science from Tsinghua University in 1970. He is currently a professor in Tsinghua University.
He is dedicated in teaching and R\&D activities in computer graphics, computer-aided design, formal verification of software, and system architecture.
He is currently the director of the School of Information Science \& Technology and the School of Software in Tsinghua University.
\end{IEEEbiography}

\begin{appendices}
	\section{  }
        \label{appendix-A}
	This Section introduces our earlier TII paper named "Semantic Learning Based Cross-Platform Binary Vulnerability Search For IoT Devices", as shown on the following page as Appendix A.
	\includepdf[pages={1-9}]{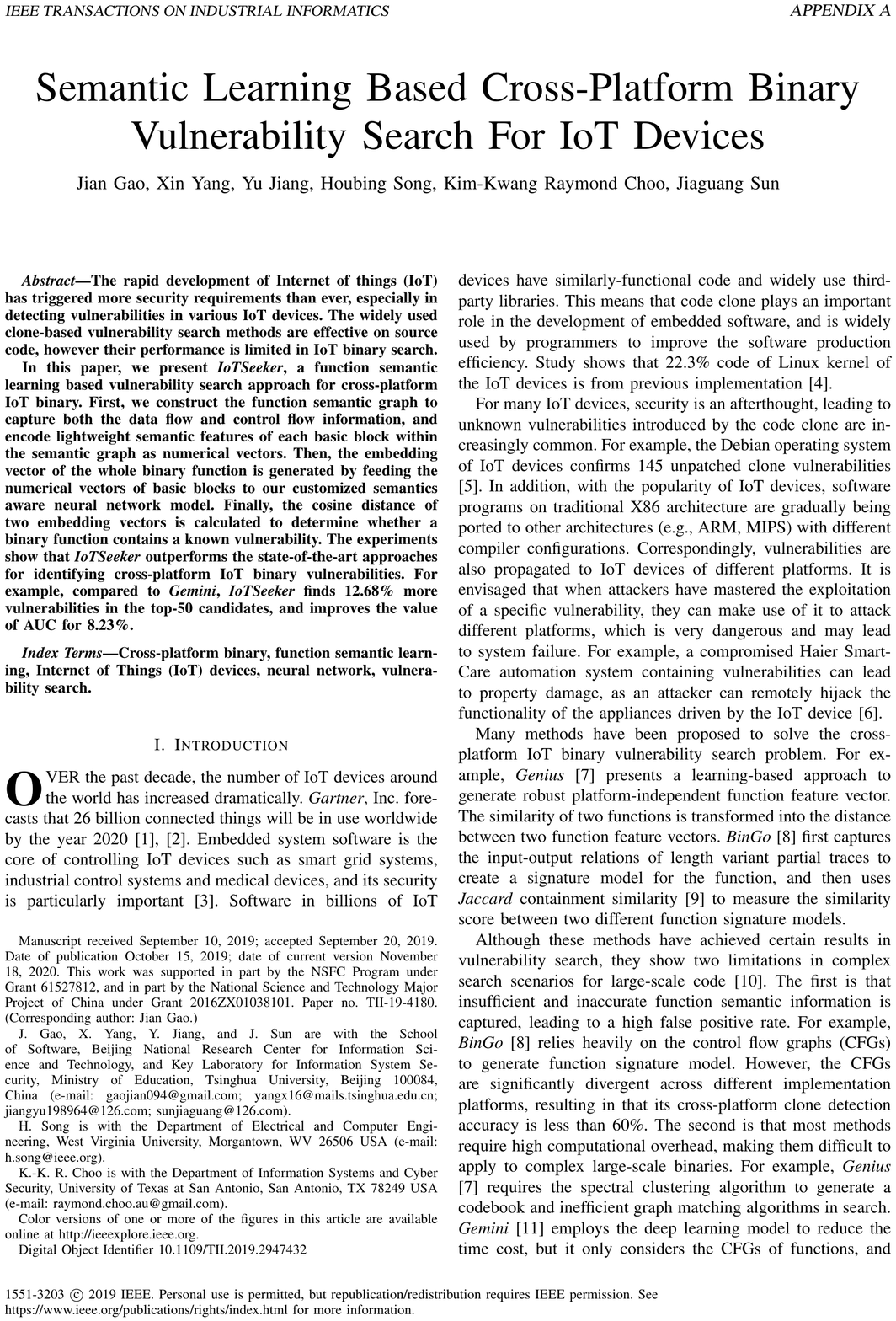}
	\section{  }
        \label{appendix-B}
	\setcounter{footnote}{0}
	This section describes the differences between the TSE paper and the TII paper. These two papers solve different research problems with different innovations, and contain different experimental designs and data charts. 

Due to both papers sharing the technical direction of code clone detection, we used several identical or similar descriptions and figures to introduce the background and machine learning. Despite the similarities in background, TSE and TII papers solve completely different problems. Specifically, the TII paper solves the problem of fast cross-architecture firmware binary vulnerability search based on machine learning; the TSE paper solves the low accuracy of pure machine learning methods and the high cost of pure semantic emulation methods. 

\begin{figure}[!htbp]
	\centering
	\includegraphics[width=3.4in]{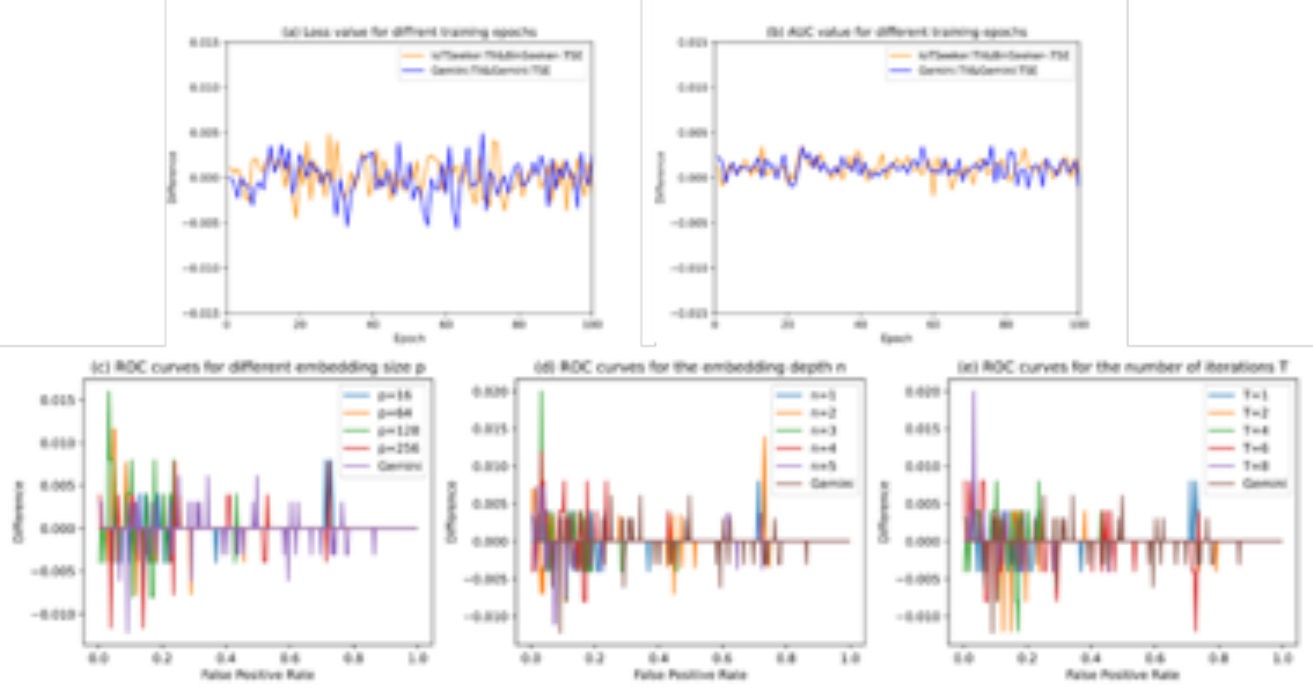}
	\caption{Comparison of data difference (first five subfigures of TSE-Fig. 7 and TII-Fig. 6).}
	\label{fig-data-difference}
\end{figure}

\begin{figure}[!htbp]
	\centering
	\includegraphics[width=3.4in]{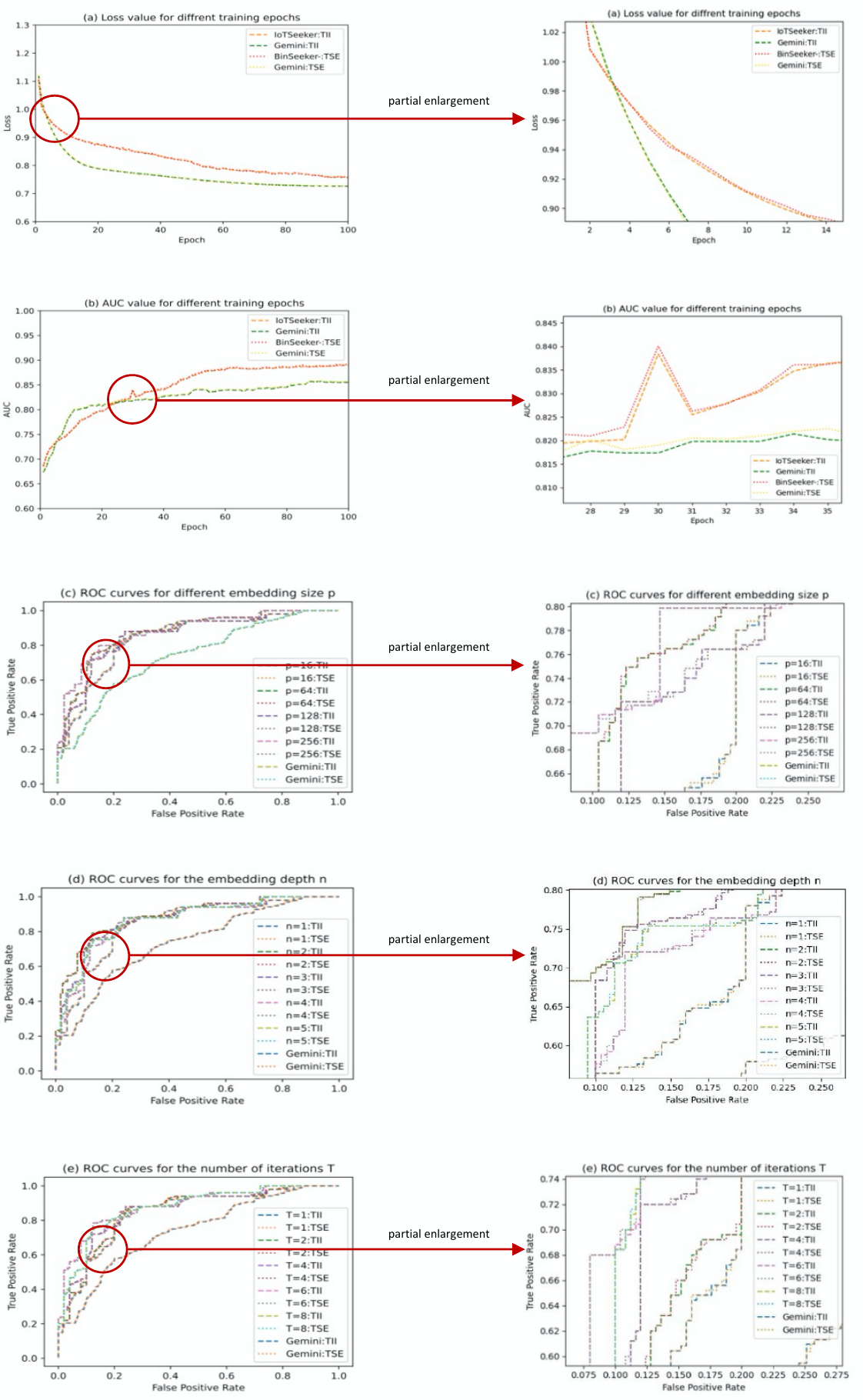}
	\caption{Overlays of the original data of the two papers (left column) and the zoomed regions (right column).}
	\label{fig-partial-enlargement}
\end{figure}

For a better understanding of these papers, we compare these papers thoroughly. In summary, they solve different problems with completely different insights. 
As a result, the experiments’ designs, data, and figures are also different. Specifically,
\begin{itemize}
	 \item \textbf{Different problems and insights.} The TII paper focuses on the detection accuracy of machine learning methods. Its technical route was completely based on machine learning. It solves the problem of fast cross-architecture firmware binary vulnerability search. Its insight is to propose a function labeled semantic flow graph and lightweight instruction features to improve the accuracy of vulnerability clone detection in the machine learning method.
	 The TSE paper focuses on balancing the accuracy of machine learning and the efficiency of semantic emulation. Its technical route integrates machine learning and semantic emulation. It solves the problem of the low accuracy of the pure machine learning method and the high cost of the pure semantic emulation method. Its insight is to propose a two-stage vulnerability clone detection method: the first stage, i.e., machine learning, achieves rapid screening of suspected vulnerability functions; the second stage, i.e., semantic emulation, achieves accurate detection of vulnerabilities. Therefore, a better balance between accuracy and efficiency is reached.
	 
	 \item  \textbf{Different experiment designs.} Due to the different insights and technical routes of TII and TSE papers, the TII paper evaluates the accuracy of machine learning methods with a three-part evaluation. The TSE paper focuses on evaluating the combined effect of semantic emulation and machine learning. The evaluation contains five parts, including the search time and accuracy improvement of semantic emulation, the effectiveness of the semantic flow graph and feature selection.
	 
	\item \textbf{Different data and figures.} Due to the limitation of processor architecture support in the semantic emulation technology used by the TSE paper, the dataset of the TSE paper is simpler. Because the base dataset is different, the models learned are different, and the evaluation data are different in turn. The main reason behind similar figures is similar evaluation metrics and analysis techniques — they are the fundamental experiment designs used across a majority of machine learning works. Furthermore, the two figures (TSE-Fig.7 and TII-Fig.6) in the hyper-parameter research section on machine learning look similar, but they are different figures based on different data sets. To illustrate the similarities and differences between the first five subfigures (a-e) in these two figures, we calculated the y-axis difference (shown in Fig.~\ref{fig-data-difference} of Appendix~\ref{appendix-B}) of the original data in the two figures. We also generated the overlays of the original figures (shown in Fig.~\ref{fig-partial-enlargement} of Appendix~\ref{appendix-B}) and zoomed to representative regions for clarity. Obviously, there are a lot of subtle fluctuations in the experimental data of the two papers. Additionally, other similar figures are used to introduce the foundation works of the two papers. Consequently, we followed the original literature’s expression and cited them correctly.
\end{itemize}

The relevant data and code of the two papers are as follows.
	\begin{itemize}
		\item TII code: https://github.com/buptsseGJ/IoTSeeker
		\item TSE code: https://github.com/buptsseGJ/BinSeeker
		\item Plotting code for the original and overlay chart, together with the original experimental data: https://github.com/buptsseGJ/ExpData
\end{itemize}

\end{appendices}

\end{document}